\documentclass[11pt,a4paper]{article}
\usepackage{amsmath,amssymb,amsthm,amsfonts}
\usepackage[dvips]{graphicx}
\usepackage{epsfig}
\usepackage{siunitx}	
\usepackage{subfigure}
\usepackage{float}
\usepackage{verbatim}
\usepackage[font=small]{caption}
\usepackage{soul}

\usepackage{color}

\title{Non-equilibrium molecular dynamics and continuum modelling of transient freezing of atomistic solids}
\author{Francesc Font\footnote{Centre de Recerca Matem\`{a}tica, Campus de Bellaterra  Edifici C, 08193 Bellaterra, Barcelona, Spain.}, William Micou\footnote{Department of Chemistry, Molecular Sciences Research Hub, Imperial College, W12 0BZ, London, United Kingdom}, Fernando Bresme\footnotemark[2] }

\def\p{\partial}

\def\({\text{\huge (}}
\def\){\text{\huge )}}
\def\bf{\textbf}
\def\]{\text{\huge ]}}
\def\[{\text{\huge [}}

\def\bf{\textbf}

\providecommand{\keywords}[1]{\textbf{\textit{Keywords:}} #1}

\newcommand{\et}{\emph{et al}.\ }
\newcommand{\bi}{\begin{itemize}}
\newcommand{\ei}{\end{itemize}}
\newcommand{\be}{\begin{equation}}
\newcommand{\ee}{\end{equation}}
\newcommand{\ba}{\begin{align}}
\newcommand{\ea}{\end{align}}

\newcommand\nc{\newcommand}
\nc\pad[2]{\frac{\p #1}{\p #2}} \nc\padd[2]{\frac{\p^2 #1}{\p
{#2}^2}} \nc\nd[2]{\frac{d #1}{d #2}} \nc\pat[2]{\frac{D #1}{D
#2}} \nc\ov{\overline} \nc\degree{^{\circ}} \nc\ord[1]{{\cal
O}(#1)} \nc\ra{\rightarrow} \nc\Ra{\Rightarrow} \nc\dint{{\mbox ~
d}}

\newcommand{\bea}{\begin{eqnarray}}
\newcommand{\eea}{\end{eqnarray}}
\newcommand{\beas}{\begin{eqnarray*}}
\newcommand{\eeas}{\end{eqnarray*}}

\begin{document}
\maketitle

\begin{abstract}
In this work we investigate the transient solidification of a Lennard-Jones liquid using non-equilibrium molecular dynamics simulations and continuum heat transfer theory. The simulations are performed in slab-shaped boxes, where a cold thermostat placed at the centre of the box drives the solidification of the liquid. Two well-defined solid fronts propagate outwards from the centre towards the ends of the box until solidification is completed. A continuum phase change model that accounts for the difference between the solid and the liquid densities is formulated to describe the evolution of the temperature and the position of the solidification front. Simulation results for a small and a large nanoscale system, of sizes $30.27$\,nm and $60.54$\,nm, are compared with the predictions of the theoretical model. Following a transient period of $\sim$20-40~ps and a displacement of the solidification front  of 1-2.5~nm we find that the simulations and the continuum theory show good agreement. We use this fact to combine the simulation and theoretical approaches to design a simple procedure to calculate the latent heat of the material. We also perform simulations of the homogeneous freezing  process, {\it i.e.} in the absence of a temperature gradient and at constant temperature, by quenching the liquid at supercooled temperatures. We demonstrate that the solidification rate of homogenous freezing is much faster than the one obtained under a thermal gradient for systems of the same size subject to the same thermostat temperature. Our study and conclusions should be of general interest to a wide range of atomistic solids.  
\end{abstract}


\section{Introduction}

Continuum heat transfer models based on Fourier law are widely used in the investigation of heat transport problems in science and engineering. These models reproduce accurately the transient cooling and heating of macroscopic systems~\cite{Incropera}. The advances in nanomaterials and nanodevices bring new challenges to characterize and quantify thermal transport in situations where Fourier law may not hold \cite{Volz16}. The classical heat diffusion equation (HDE), in combination with non-equilibrium molecular dynamics (NEMD) simulation methods, has been used to investigate transient cooling of nanoscale structures and to obtain transport coefficients such as the thermal conductance \cite{B817666C,C6CP06403E,rajabpour} or the thermal conductivity \cite{Mul97,Tretiakov2004}. Previous studies revealed deviations between the transient molecular dynamics simulations and HDE predictions \cite{Volz96,LIU2008,FontBresme2018}. These deviations are observed at very short times (at picosecond and sub-picosecond time scales), at conditions that deviate significantly from the equilibrium~\cite{Volz96} or due to the interplay of activation processes~\cite{FontBresme2018}. 

Continuum heat transfer models have been extended to investigate thermal transport processes involving phase changes, such as melting and freezing \cite{Crank,Gupta,Alex93}. These models typically involve the solution of the HDE in space domains whose geometry is evolving with time. Classical phase change models have been employed to describe the melting of nanoparticles, nanowires and nanoslabs \cite{FontBresme2018,Wu09,Font2013,Font2015,Back14,Florio2016}. However, the application of the classical HDE to nanoscale materials might be limited by deviations from the diffusive regime, due to the phonon mean free path becoming comparable to or larger than the characteristic lengthscale of the material \cite{chen2}. Efforts have been made to formulate alternative heat transport equations that account for such deviations and allow the description from macro to nanoscale systems, {\it e.g.} see \cite{Alvarez07,Vazquez,Guo18}. Recently, some of these approaches have been extended to formulate theoretical models that include phase change \cite{Font2018,HENNESSY20181,CALVOSCHWARZWALDER2020106210}. However, the limits of applicability of the HDE or the classical models describing phase change are yet to be determined accurately, partially due to the difficulty in designing experiments at the nanoscale, which would serve as tests of the theory. Instead, non-equilibrium molecular dynamics provides exact results of well defined models, and therefore a suitable reference to test the accuracy of the continuum theories. 

Intense illumination of nanocrystals using high power lasers can trigger a cascade of melting and freezing processes~\cite{Clark7444,Kirschner18,Kirschner,Thompson2012,Font2017}, which are important when considering the characterization of nanomaterials using {\it e.g.} x-ray diffraction probe experiments or in nanoparticle self-assembly induced by dewetting of thin nanoscale metallic films \cite{kondic2020}. In \cite{Eswaramoorthy1437} the composition and evolution of solid-liquid interfaces during the solidification of partially molten aluminium-based micro/nanocrystals alloys were analysed using energy-dispersive x-ray spectroscopy. We investigated recently~\cite{FontBresme2018} the transient melting problem of a simple atomic solid using both the classical phase change model and transient non-equilibrium molecular dynamics simulations. The classical model reproduces very accurately the dynamics of melting of small crystals. While the non-equilibrium simulations validated the Stefan condition, namely, the melting proceeds with an interfacial temperature equal to the thermodynamic melting temperature, they also revealed deviations at short times, $< 100$~ps. The differences between the molecular dynamics (MD) simulations and the theory were interpreted in terms of the activated nature of melting, and the time required to generate a nucleus of liquid inside a cold solid. 

Here, we investigate the inverse problem, namely the transient solidification of a liquid. We compare the solutions for the dynamics of freezing obtained by the continuum phase change model with the predictions of non-equilibrium MD simulations. We have chosen the same atomic fluid studied in our transient melting work. The details of the MD simulations and the phase change model are presented in Sections 2.1 and 2.2, respectively. In Sections 3.1 and 3.2, we present results that support the accuracy of the HDE at reproducing the temperature profiles from very small times and until the completion of freezing. The simulations also support the assumptions in the theory, namely, the temperature continuity at the solid-liquid interface and the energy balance at the interface (the Stefan condition). In section 3.3 we combine the theory and simulation results and propose a simple procedure to calculate the latent heat of solidification. In section 3.4 we compare the crystallisation rate of the transient process in the presence of a thermal gradient with the rate obtained for the freezing process at homogeneous temperature. This topic is of interest given the fast crystallisation rates observed in liquid under homogeneous supercooling conditions, as well as the proposed lack of activation energy for the transient crystallisation processes in atomistic solids~\cite{Broughton1982}. Finally, in Section 4 we present our conclusions. 

\section{Methods}

\subsection{Non-equilibrium simulations}

\begin{figure}[ht]
\includegraphics[scale=.5,angle=-90]{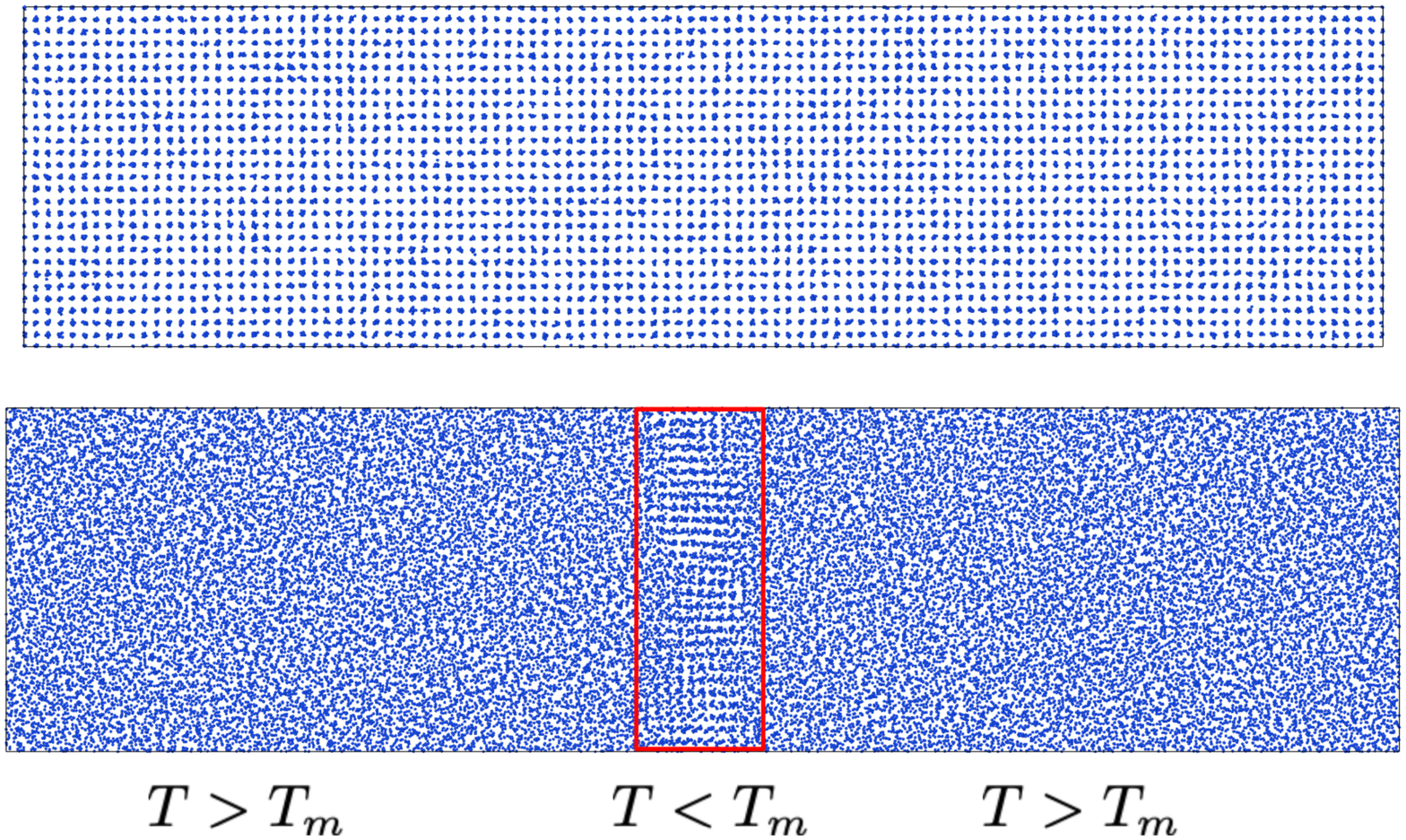} \caption{Snapshots showing a pre-equilibrated system at $P=3.893$. (Top) Solid at $T^*=0.8$ equilibrated using an isotropic barostat. (Bottom) The simulated system after 2$\times 10^4$ time-steps, showing the solid seed region in the center at $T^*=0.8$ and two liquid regions at $T^*=1.4 > T_f^* = 1.0$. The system contains 43904 atoms, corresponding to 14$\times$14$\times$56 face-centered cubic (fcc) unit cells. The red rectangle (thickness 8$\sigma$) in the middle of the simulation cell signals the location of the cold thermostat. 
}
\label{fig1}
\end{figure}

The computer simulations were performed using transient non-equilibrium molecular dynamics following the approach discussed in reference~\cite{FontBresme2018}.  The initial system was obtained from a pre-equilibrated crystal simulated at constant pressure and temperature \cite{npt}. These simulations were performed using an isotropic barostat. Since freezing from a melt is an activated process, the nucleation might not take place instantaneously. For the opposite process---nucleation of melting in a solid---we reported time delays between 5\,ps and 80\,ps for thermostat temperatures between 40\% and 25\% above the melting temperature, respectively, for the same Lenard-Jones system investigated here~\cite{FontBresme2018}. Also, it might be difficult to control the symmetry of the face growing in the thermal field due to stresses arising in the periodic simulation box. Hence, we decided to perform an additional equilibration process. We defined a region in the middle of the simulation cell that was thermostatted (using simple rescaling of the velocities) at a temperature below the thermodynamic freezing temperature, $T_f$, while the rest of the system was heated at $T>T_f$. These simulations were performed using an anisotropic barostat, by coupling the cell vectors parallel to the liquid-solid interface, $L_x$ and $L_y$,  and uncoupling the vector in the direction perpendicular to the interface plane, $L_z$. The final system consists of a solid slab, which acts as a seed for the nucleation of the surrounding liquid, during the transient freezing simulations. We show in Figure \ref{fig1} one snapshot  of the systems employed to generate the initial configuration to study the transient freezing process.

We performed simulations of the truncated and shifted Lennard-Jones potential,
\begin{equation}\label{tslj}
\begin{aligned}
U(r) &= 4 \epsilon \left[ \left(\frac{\sigma}{r} \right)^{12} -  \left(\frac{\sigma}{r} \right)^{6} \right ] & \mbox{for}~r~\leq~r_c\,,\\
U(r) &= 0&~\mbox{for}~r~>~r_c\,,
\end{aligned}
\end{equation}
where $\sigma$ and  $\epsilon$ define the atom diameter and the atom-atom interaction strength, respectively, and the cutoff $r_c$ was set to 6$\sigma$. This cutoff provides very similar results to those obtained with the full potential. The solid-liquid equilibrium of this model has been investigated extensively, and an accurate equation of state was reported~\cite{mastny}. Furthermore, the potential has been used to simulate thermodynamic and elastic properties of face centered metals, predicting accurate results \cite{heinz}.  We used it in our previous work to investigate the inverse problem discussed in this work, namely transient melting \cite{FontBresme2018}, hence, it provides a good reference to study transient freezing. 
The scales $\sigma$ and $\varepsilon$ were used to define reduced units: $T^*=k_B T / \varepsilon$ and $\rho^*=\rho_n \sigma^3$, where $T$ and $\rho_n$ are the temperature in Kelvin and the number density in particle per m$^3$, respectively, and $k_B$ the Boltzmann constant. Conversion between reduced units and their corresponding dimensional values can be obtained multiplying the reduced variable by the corresponding Lennard-Jones time, length or temperature scale, for example for argon: $\varepsilon=1.65\times 10^{-21}$~J ($\varepsilon/k_B$=119.48~K), $\tau=2.17\times10^{-12}$\,s, $\sigma=3.405\times10^{-10}$\,m. 

Hereafter, we employ standard reduced units using $\sigma$ and $\varepsilon$ length and energy scales, to report our results. The equations of motion were integrated using the velocity Verlet algorithm with a timestep $\delta t^* = 0.002$. We investigated systems consisting of 14 unit cells in the $x$ and $y$ directions, and 56 or 112 unit cells in the $z$ direction. Typical system sizes ranged between 43904 and 87808 atoms. The cold thermostat in the middle of the box acting as heat sink for the freezing process was set to either  $T^*=0.6$ or $T^*=0.8$, while the initial temperature of the heated liquid surrounding the solid slab in the center of the simulation box was set to $T^*=1.3$ or $T^*=1.4$, respectively. Following our previous work \cite{FontBresme2018}, we find that the transient process involves a change in the average temperature of the simulation cell, and therefore a change in the system pressure if the simulation is performed at constant volume.  To maintain the system at a pressure compatible with coexistence conditions, we performed the simulations using the Berendensen barostat (the time relaxation constant was set to 2 in reduced units of time), with the $L_x$ and $L_y$ simulation cell vectors coupled and the $L_z$ varying independently. 
The presence of the liquid-solid interface should have a minor impact on the use of the barostat (see ref. \cite{FontBresme2018}). The transient freezing process was followed along the \{1,0,0\} face, by setting the initial solid seed  (see central region in Figure~\ref{fig1}) in the appropriate orientation.

The time-dependent properties were calculated  in the direction normal to the interface, $z$, by dividing the simulation box in bins of thickness $\sigma/2$. The local temperature was computed using the equipartition principle, 
\begin{equation}
\label{temperature}
\frac{1}{ k_B T(z)} = \left< \frac{3 N} {\sum_{i=1}^{N \in z} \frac{{\bf p}_i^2}{m_i}} \right >\,,
\end{equation}
\noindent
where ${\bf p}_i$ is the momentum of particle $i$ and the sum runs over the particles in bin $z$. We also computed the $Q_6$ order parameter, which is given by setting $l=6$ in the definition of the Steinhardt's order parameters \cite {PhysRevB.28.784}, 
\begin{equation}
Q_l = \sqrt{\frac{4\pi}{2l+1} \sum_{m=-l}^{m=l} Y_{lm} Y_{lm}^*}\,, 
\end{equation}
\noindent
where $Y_{lm}$ are the spherical harmonic order parameters. The $Q_6$ is rotationally invariant and defines the local bond orientational order of each atom in the system. Steinhardt's order parameters were introduced by Steinhardt et al. to characterize the local orientational order in atomic structures, and adopt well defined values for crystals with specific symmetries \cite {PhysRevB.28.784}. 
We selected 12 as the number of nearest neighbours, and a cutoff for neighbour search of 2$\sigma$. Averages for temperature, density and order parameter profiles were obtained over seven independent transient simulations. Each simulation lasted 10$^6$ time steps, to ensure the complete freezing of the system. 

We also performed simulations of homogeneous freezing. {\it i.e.} at constant temperature,  by quenching the temperature of the entire system to supercooled temperatures $T^*=0.8$ and $T^*=0.85$ below the freezing temperature, $T_f^* = 1$, to compare the crystallisation rates obtained with or without thermal gradients.  We tried different thermostatting strategies, Nose\'e-Hoover (with the Parrinello-Rahman barostat~\cite{parrinello1981}), v-rescale~\cite{bussi2007} (with the Berendsen barostat) or Langevin (with the Berendsen barostat) ~\cite{schneider1978}, obtaining similar rates for the systems studied here. All the trajectories were generated with LAMMPS~\cite{PLIMPTON19951}.

\subsection{Continuum solidification model} 

One of the objectives of our work is to investigate the limits of applicability of classical heat transfer theory when attempting to describe phase change at the nanoscale. In this section, we formulate a one-dimensional phase change model based on continuum heat transfer theory that takes into account the different densities between the solid and the liquid, which is a feature typically neglected in phase change models. The model consists on HDEs for the solid and the liquid and an equation for the evolution of the solidification front, namely the Stefan condition.   

\begin{figure}[ht]
\includegraphics[width=\textwidth]{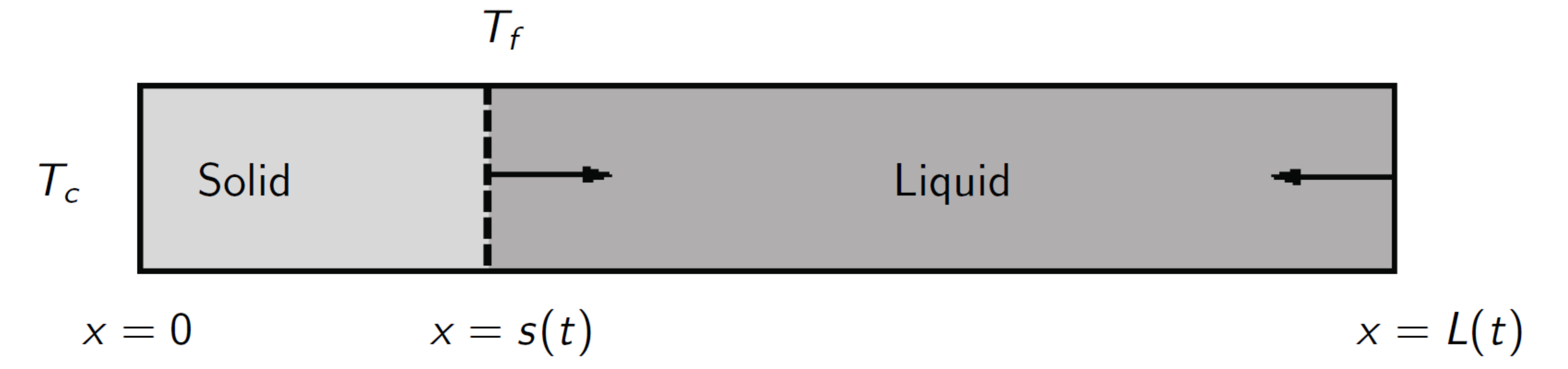} 
\caption{Illustration of the continuum phase change model. The reference $x=0$ corresponds to the left/right edge of the thermostat in the MD simulations and  
	$s(t)$ the displacement of the left/right solidification front relative to the edge of the thermostat.}
\label{illusatrationmodel}
\end{figure}

We consider an idealised one-dimensional liquid initially at temperature $T_0$, occupying the space $0\leq x\leq L$. Suddenly, the temperature is lowered to $T_c$ below the freezing point $T_f$ on the edge $x=0$ and the liquid starts to solidify. The newly created solid phase will start to grow, occupying the space $0<x<s(t)$, where $s(t)$ represents the position of the solidification front (solid-liquid interface). Due to the density difference between the liquid and the solid, the total size of the system will decrease, {\it i.e.} $L=L(t)$. Hence, the end of the domain is allowed to adjust and the liquid phase will occupy the space $s(t)<x<L(t)$. An illustration of the model is shown in Figure~\ref{illusatrationmodel}. The reference $x=0$ (see Figure~\ref{illusatrationmodel}) in the continuum model corresponds to the edge of the thermostat (either left or right side of the red square in Figure~\ref{fig1}) from the MD simulations. Therefore, $s(t)$ indicates the displacement of the solid-liquid interface relative to the edge of the thermostat. We note that the model can reproduce the propagation of either the left or right solidification fronts occurring in the simulations, since they are symmetric. Since the solid-liquid interface at $t=0$ is located on the edge of the thermostat we have that $s(t=0)=0$.

The temperatures of the solid and liquid phases, $T_s(x,t)$ and $T_l(x,t)$, are described by the heat diffusion equations
\begin{align}
\rho_s c_s \frac{\partial T_s}{\partial t} &= k_s \frac{\partial^2 T_s}{\partial x^2} \qquad \mbox{on} \qquad 0<x<s(t) \,,\label{eq1}\\
\rho_l c_l \left(\frac{\partial T_l}{\partial t}+v_l\frac{\partial T_l}{\partial x}\right) &= k_l \frac{\partial^2 T_l}{\partial x^2} \qquad \mbox{on} \qquad s(t)<x< L(t) \,,\label{eq2}
\end{align}
where $\rho_{j}$ is the density, $c_j$ the specific heat, $k_{j}$ the thermal conductivity and the subscripts $j=$ `$s$' or `$l$' indicate  solid or liquid, respectively. The velocity $v_l=dL/dt$ is the velocity in the fluid due to the change in density, in our case mass conservation gives $v_l=-(\rho_s/\rho_l-1)ds/dt$ (see, \cite{Alex93,Font2015}). The temperatures of the solid and the liquid are subject to the boundary conditions 
\begin{align}\label{eq3}
T_s(0,t) = T_c\,,\qquad T_s(s(t),t) = T_l(s(t),t) = T_f\,, \qquad  \left.\frac{\partial T_l}{\partial x}\right|_{x=L(t)} = 0\,,  
\end{align}
where $T_c$ represents the temperature of the thermostat which drives the solidification. We have assumed a no-flux boundary condition at $x=L(t)$ to reproduce the equivalent situation described by the periodic boundary conditions in our simulations. Finally, at the solidification front we have the Stefan condition
\begin{align}\label{eq4}
\rho_s\,\left[\Delta H_f+\frac{1}{2}\left(\frac{\rho_s^2}{\rho_l^2}-1\right)\left(\frac{ds}{dt}\right)^2\right]\,\frac{ds}{dt} = k_s \left.\frac{\partial T_s}{\partial x}\right|_{x=s(t)} - k_l \left.\frac{\partial T_l}{\partial x}\right|_{x=s(t)}\,, 
\end{align}
where $\Delta H_f$ is the latent heat of solidification. The term $\propto(ds/dt)^2$ in \eqref{eq4} is the kinetic energy contribution at the interface due to the bulk liquid motion \cite{Alex93,Font2015,MYERS2020118975}. The system is closed with the initial conditions 
\begin{align}\label{ic}
T_l(x,0)=T_0\,,\qquad s(0)=0\,.
\end{align}

The moving boundary problem \eqref{eq1}-\eqref{ic} does not have an analytical solution and must be solved numerically. Due to the inconvenience of solving partial differential equations in moving domains, we choose to transform the domains of the solid and the liquid phase into fixed unit domains by introducing the change of variables $\xi = x/s(t)$ and $\eta = (x-s(t))/(L(t)-s(t))$, which transform $x\in[0,s(t)]$ and $x\in[s(t),L(t)]$ into $\xi\in[0,1]$ and $\eta\in[0,1]$, respectively. The resulting equations are solved numerically using a finite differences semi-implicit backward Euler scheme, which we implement in Matlab. This is a standard procedure for these types of problems and details can be found in \cite{Font2015,Font2018}. 

A common simplification of the model can be obtained by noting that $\rho_s\approx\rho_l$, which allows to neglect the advection term in \eqref{eq2} (since $(\rho_s/\rho_l)-1\approx 0$) and the term $\propto (ds/dt)^2$ in \eqref{eq4} (since $(\rho_s/\rho_l)^2-1\approx 0$). By taking this assumption we also prevent the system from shrinking due to the solid-liquid density change. This reduction was used in our previous work \cite{FontBresme2018}. In the results section we shall refer to \textit{Case $\rho_s\neq\rho_l$} as the full version of the model \eqref{eq1}-\eqref{ic} and to \textit{Case $\rho_s\approx\rho_l$} as the version that neglects the density variation terms. 

Finally, the most classical solution form of the model is obtained assuming $\rho_l \approx \rho_s$ and taking $L\rightarrow\infty$ ({\it i.e.}, considering the system to be semi-infinite). In this case, the boundary condition at $x=L$ can be substituted by $\left.T_l\right|_{x\rightarrow\infty}= T_0$ and the problem has the exact solution 
\begin{align}\label{}
T_s(x,t) &= T_c + (T_f-T_c)\frac{\text{erf}\left(x/2\sqrt{\alpha_s t}\right)}{\text{erf}\left(\lambda\right)}\,,\label{eq5}\\
T_l(x,t) &= T_0 -(T_0-T_f)\frac{\text{erfc}\left(x/2\sqrt{\alpha_l t}\right)}{\text{erfc}\left(\lambda\sqrt{\alpha_s/\alpha_l}\right)}\,,\label{eq6}
\end{align}
where $\alpha_j=k_j/\rho_j c_j$ is the thermal diffusivity, with the solidification front propagating according to 
\begin{align}\label{eq7}
s(t) = 2\lambda \sqrt{\alpha_s\, t}\,.
\end{align}
The dimensionless parameter $\lambda$ is the solution of the transcendental equation 
\begin{align}\label{eq8}
\frac{1}{\beta_s\exp(\lambda^2)\text{erf}\left(\lambda\right)} - \frac{1}{\beta_l\sqrt{\alpha_s/\alpha_l}\exp(\alpha_s\lambda^2/\alpha_l)\text{erfc}\left(\lambda\sqrt{\alpha_s/\alpha_l}\right)} = \sqrt{\pi}\lambda\,, \end{align}
where $\beta_s=\Delta H_f/c_s(T_f-T_c)$ and $\beta_l = \Delta H_f/c_l(T_0-T_f)$ are the Stefan numbers for the solid and liquid phase, respectively. In the next section we will refer to these equations as the Neumann solution or \textit{Case $\rho_s\approx\rho_l$, $L\rightarrow \infty$}. The thermal properties for the solid and the liquid are taken from our previous  work \cite{FontBresme2018} and are summarised in Table~\ref{Table1}. 

\begin{table}
\begin{center}
\begin{tabular}{cccc}
\hline\hline
Phase & $\rho_i$ [kg/m$^3$] & $c_i$ [J/(kg$\cdot$K)] & $k_i$ [W/(m$\cdot$K)]  \\
\hline
Solid ($i=s$)  & 1690.0 & 863.62 & 0.234 \\
Liquid ($i=l$) & 1543.5 & 921.7  & 0.173 \\
\hline\hline
\end{tabular} 
\caption{Solid and liquid properties of the Lennard-Jones system simulated in this work. The enthalpy of solidification is $\Delta H_f = 32492$\,J/kg and the freezing temperature $T_f=119.45$\,K. }
\label{Table1}
\end{center}
\end{table}

\section{Results and Discussion}

In the first part of this section we will present the results obtained with the three systems simulated in this work; two small systems of size 14$\times$14$\times$56 that solidify under different cooling conditions, one at $T_c^*=0.8$ and $T_0^*=1.4$ and the other at $T_c^*=0.6$ and $T_0^*=1.3$, and a large system of size 14$\times$14$\times$112 that solidifies at the conditions $T_c^*=0.8$ and $T_0^*=1.4$. In section~\ref{3.2} the simulated temperatures and the  position of the front are compared with the solutions of the theoretical model \eqref{eq1}-\eqref{eq4}. We will present a set of three different theoretical solutions that depend on the assumptions taken to simplify the model. The analytical solution follows from assuming $L\rightarrow\infty$ and $\rho_s\approx\rho_l$. An intermediate solution is obtained by using $\rho_s\approx\rho_l$ but letting $L$ to be finite and equal to the initial size of the system. The last and most general solution is obtained by solving the full system numerically, which accounts for the fact that $\rho_s\neq\rho_l$ and allows the system to continuously shrink from its initial size $L_0$ to its final size $(1-\rho_l/\rho_s)L_0$. Then, in section~\ref{3.3} we will explain how the MD simulations and the theoretical model can be combined to calculate the  latent heat of freezing.  

In section~\ref{3.4} we present simulation results for  the solidification of a system of size 14$\times$14$\times$56 without a temperature gradient, obtained by supercooling the liquid at a homogeneous temperature. These simulations are compared with the results obtained in section~\ref{3.2}  to assess the differences  in solidification rates with homogeneous cooling or with temperature gradients.  

\subsection{Density, temperature and order parameter evolution from MD simulations}\label{3.1}

\begin{figure}[ht]
\begin{center}
\includegraphics[scale=.40,angle=-90]{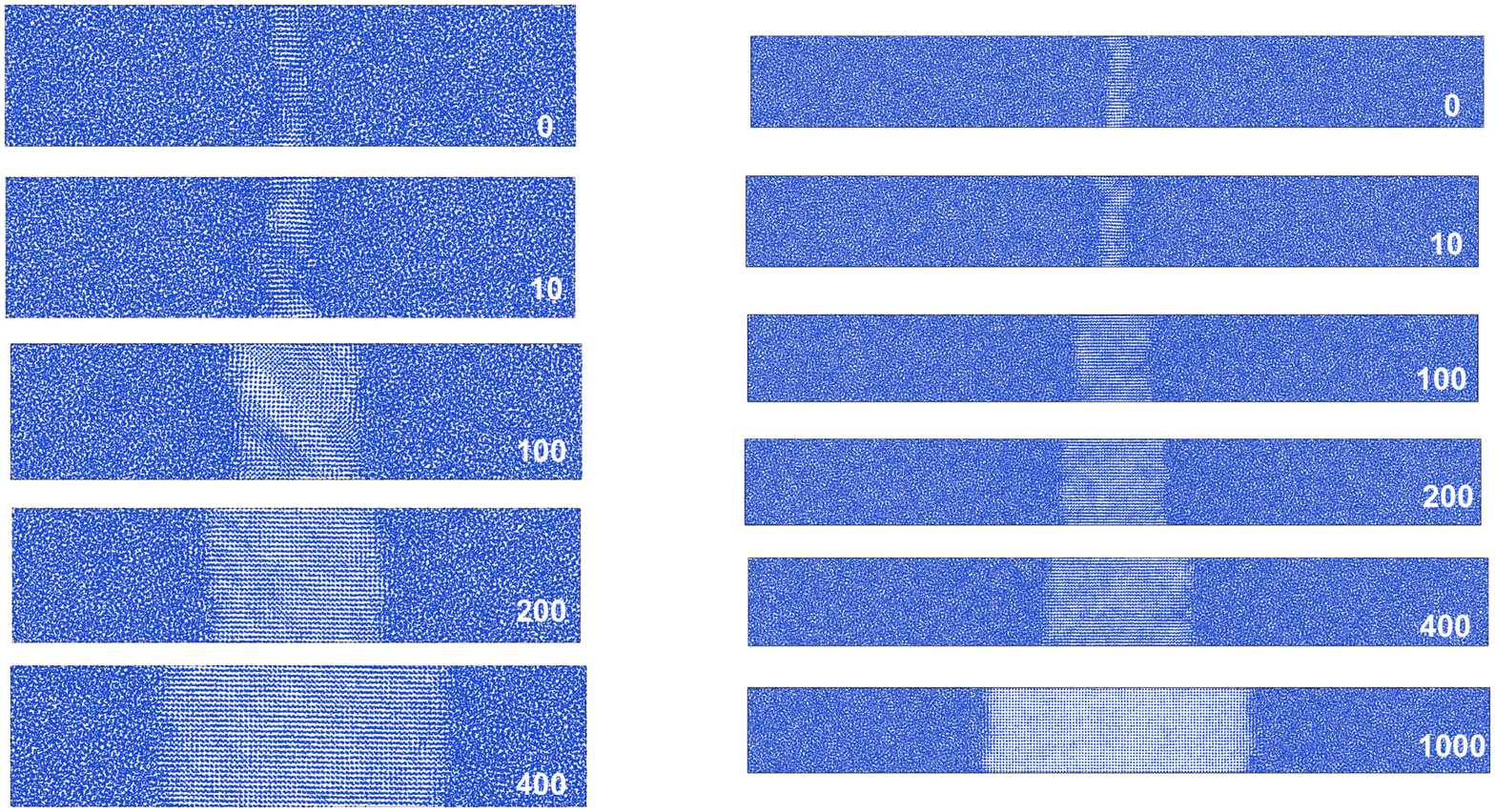} 
\end{center}
\caption{Snapshots of the transient freezing process for a system with the liquid pre-equilibrated at $T^*=1.4$ and the solid region in the center of the simulation cell at $T^*=0.8$. The snapshots illustrate the freezing process for systems consisting of 14$\times$14$\times$56 (left) and  14$\times$14$\times$112 (right) unit cells. The numbers indicate the time in Lennard-Jones units.}
\label{fig2}
\end{figure}

\begin{figure}[!h]
\begin{center}
\includegraphics[scale=.40,angle=-90]{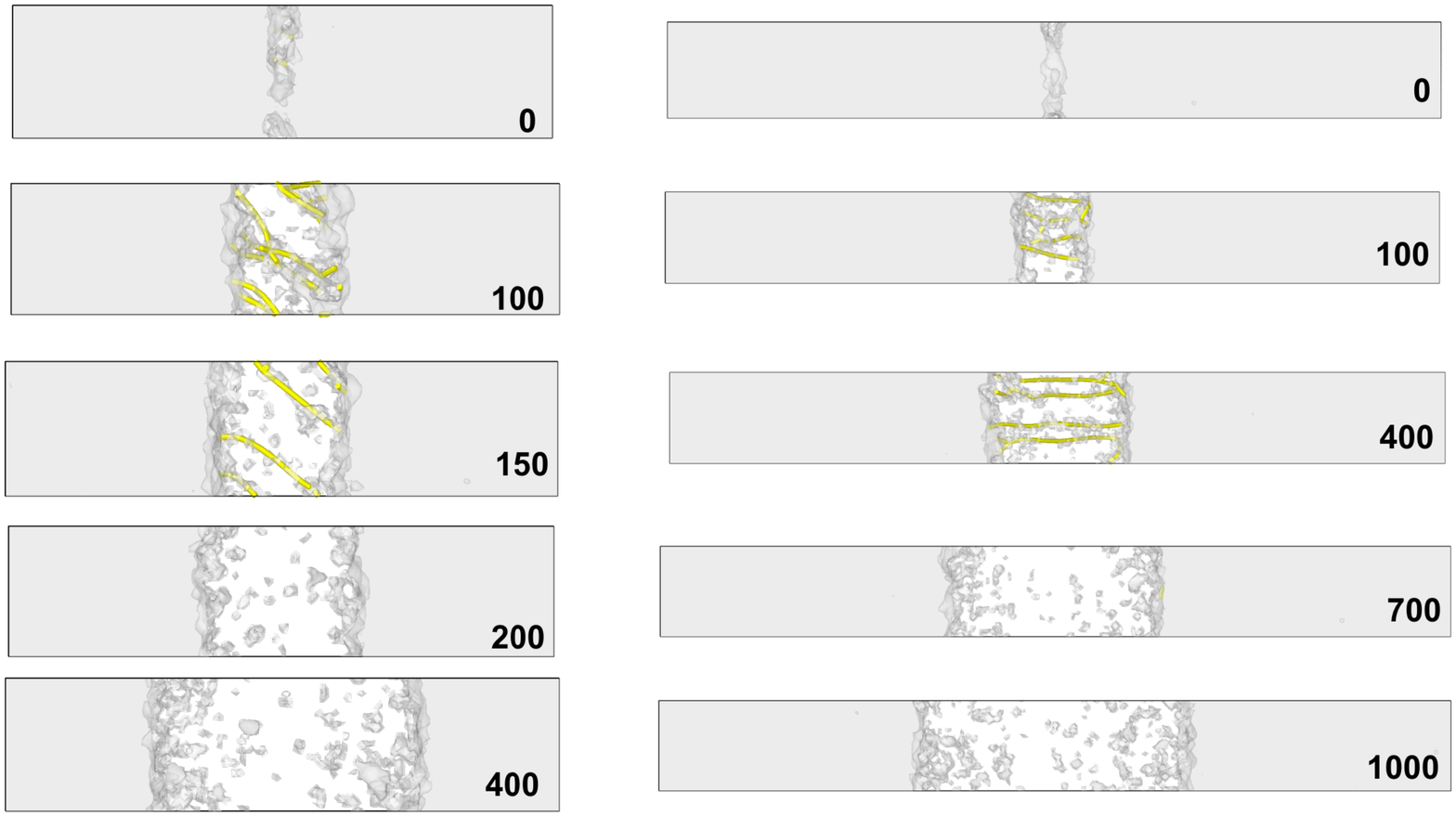}
\end{center}
\caption{Dislocation analysis of the solid structures formed during the freezing process, for systems of different sizes, 14$\times$14$\times$56 (left) and  14$\times$14$\times$112 (right) unit cells, with the liquid initially equilibrated at $T^*=1.4$ and the solid region in the center of the simulation cell at $T^*=0.8$. The grey area and grey surface indicates regions that do not conform to the fcc structure. White background indicates fcc structure. The cylindrical tubes (yellow) represent the dislocations formed in the crystal. The numbers indicate the time in Lennard-Jones units.}
\label{fig-dislocation}
\end{figure}

We show in Figure~\ref{fig2}, a sequence of snapshots illustrating the time evolution of the freezing front for the small (left) and the large (right) systems. The snapshots show that our simulation approach generates crystals with the correct symmetry, hence supporting our approach of using a well-defined seed with face-centered cubic (fcc) symmetry to promote the nucleation of the solid.  The structure of the crystals formed during  the transient freezing process was analysed using the dislocation extraction algorithm  \cite{Stukowski_2012,ovitoref} (see Figure~\ref{fig-dislocation}). At short times, the crystal  develops some dislocations (see cylindrical tubes in Figure~\ref{fig-dislocation}), but these reduce significantly over time before disappearing  in about 200-400 time units (400-800 ps).

\begin{figure}[!h]
\begin{center}
\begin{tabular}{c c}
\includegraphics[scale=.450,angle=0]{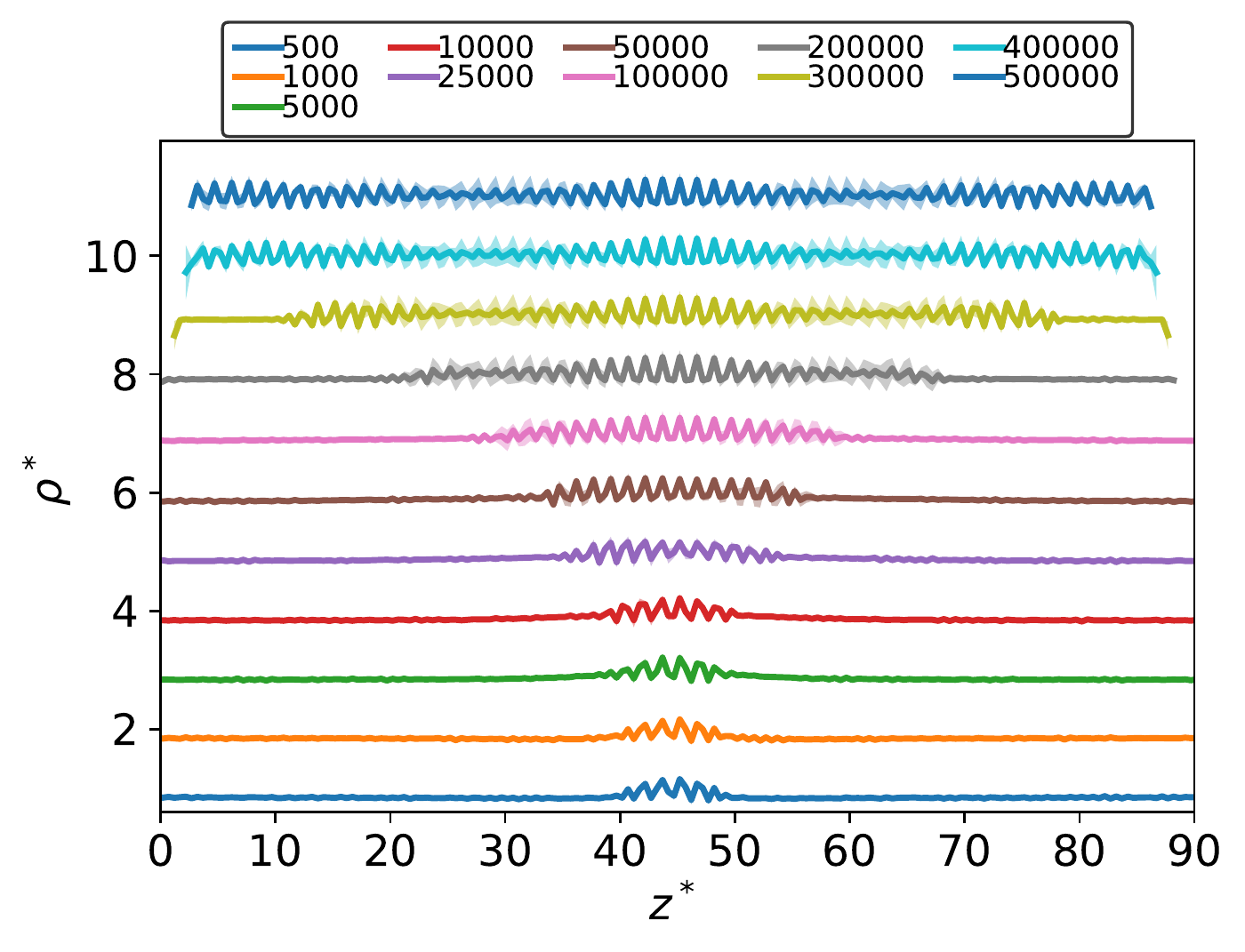} &  \includegraphics[scale=.450,angle=0]{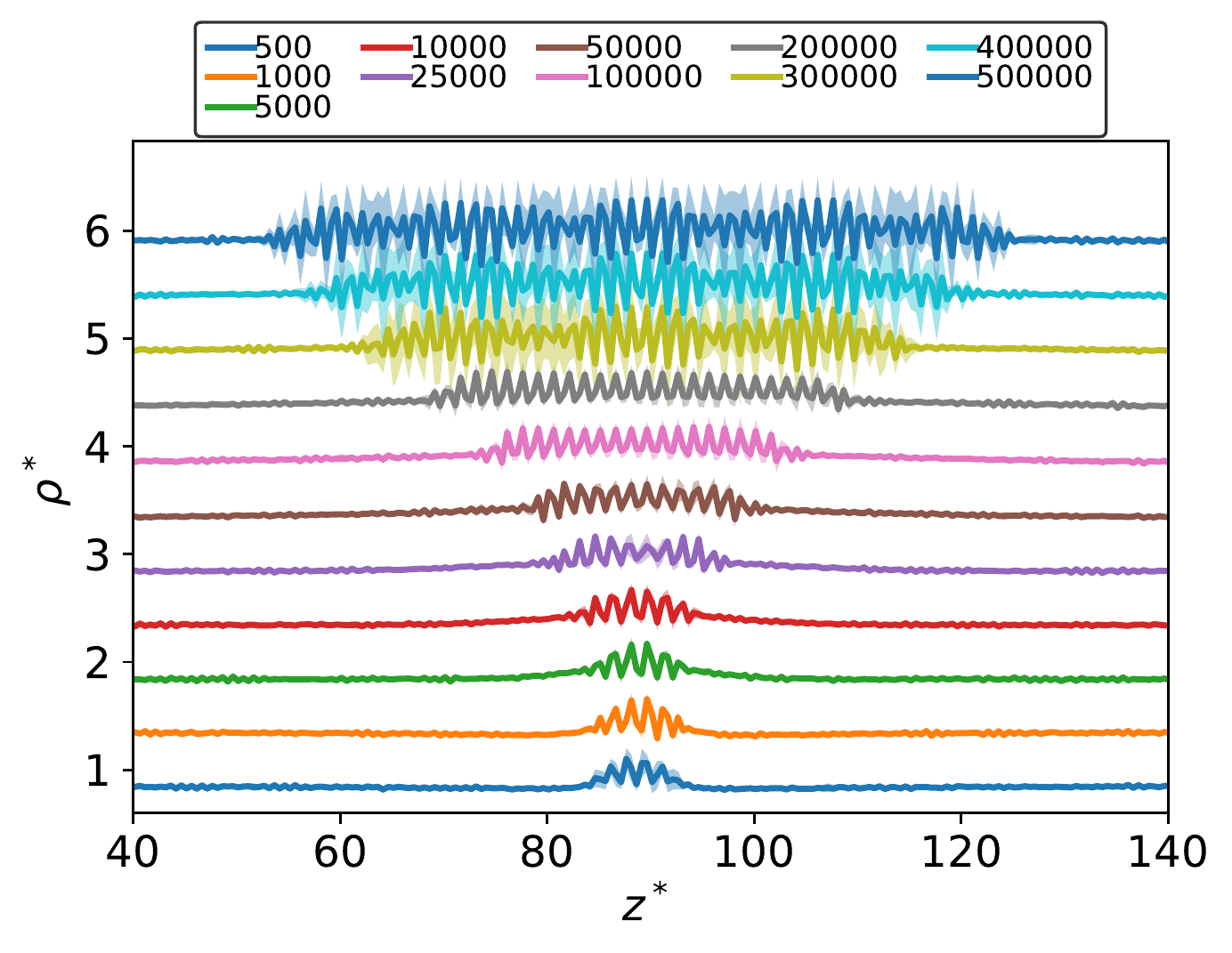}  \\
\includegraphics[scale=.450,angle=0]{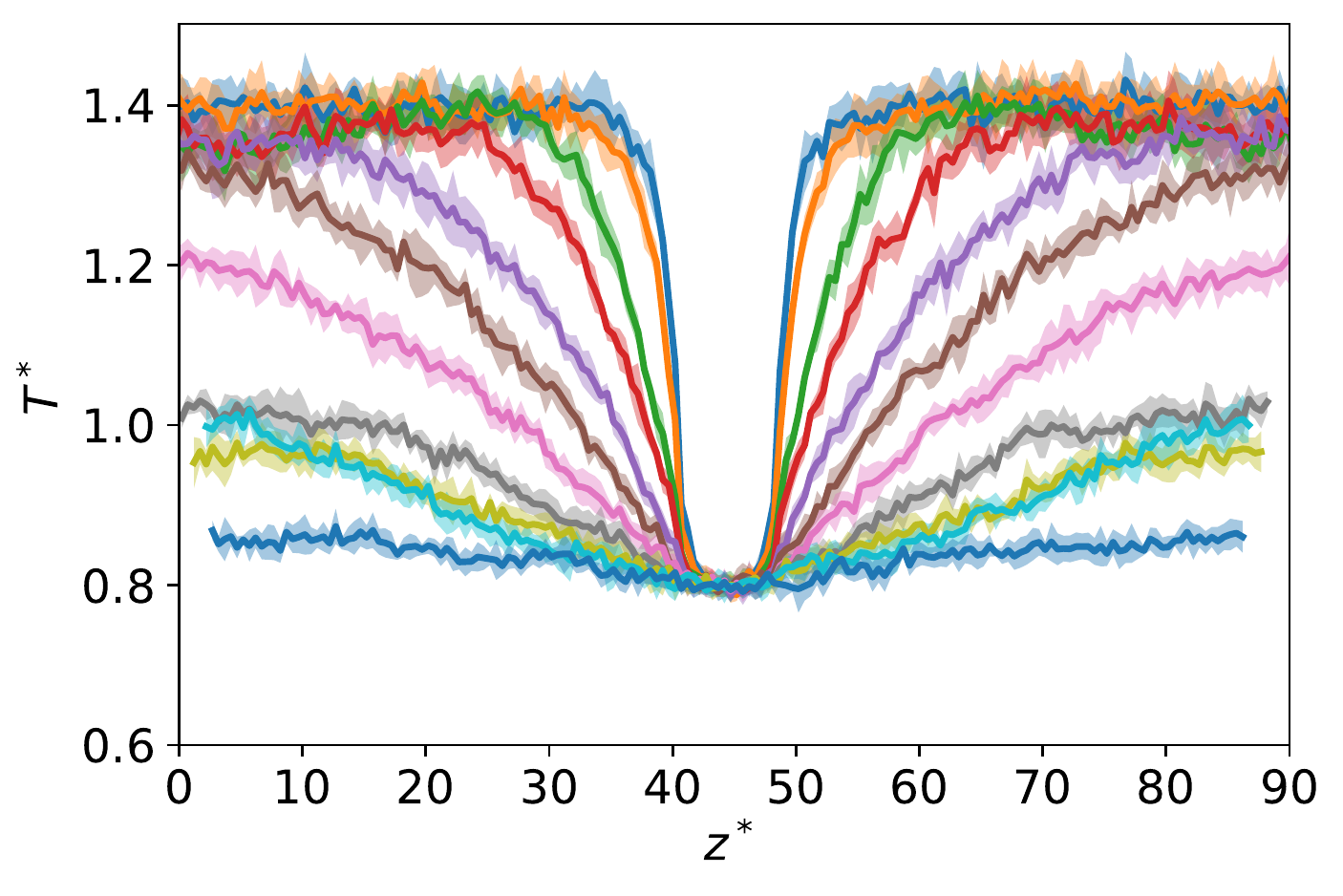} &  \includegraphics[scale=.450,angle=0]{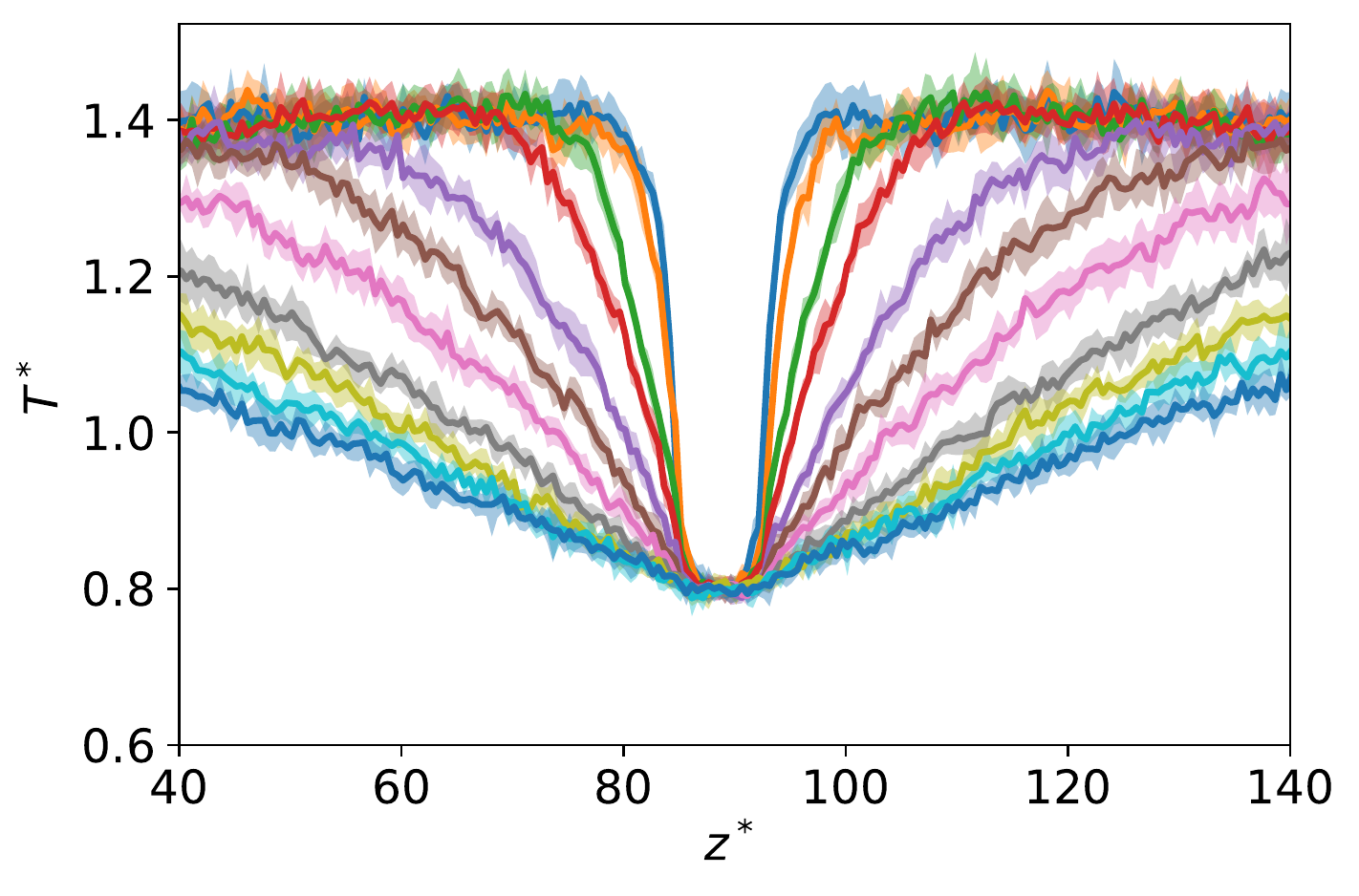} \\
\includegraphics[scale=.450,angle=0]{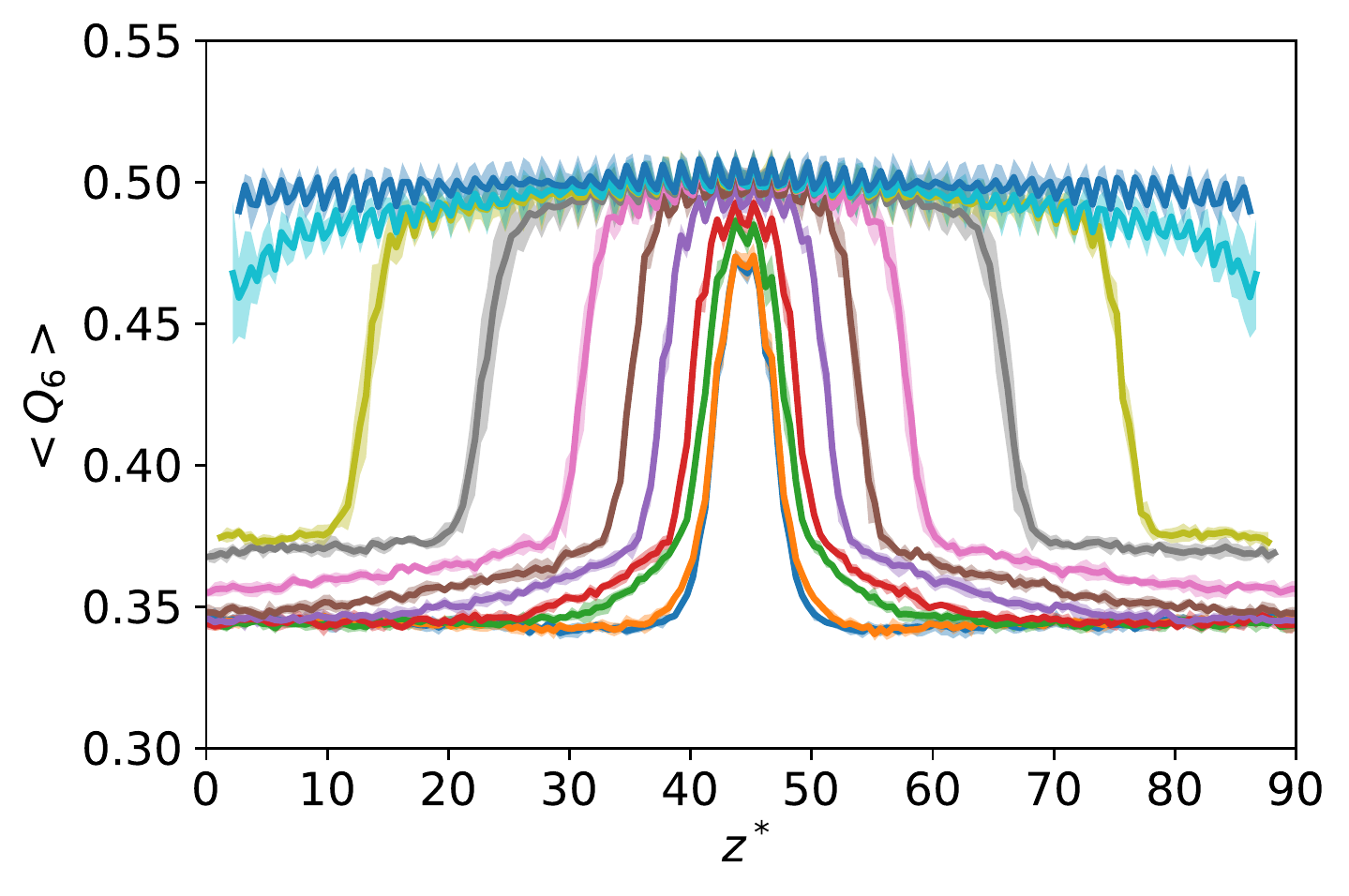} & 
\includegraphics[scale=.450,angle=0]{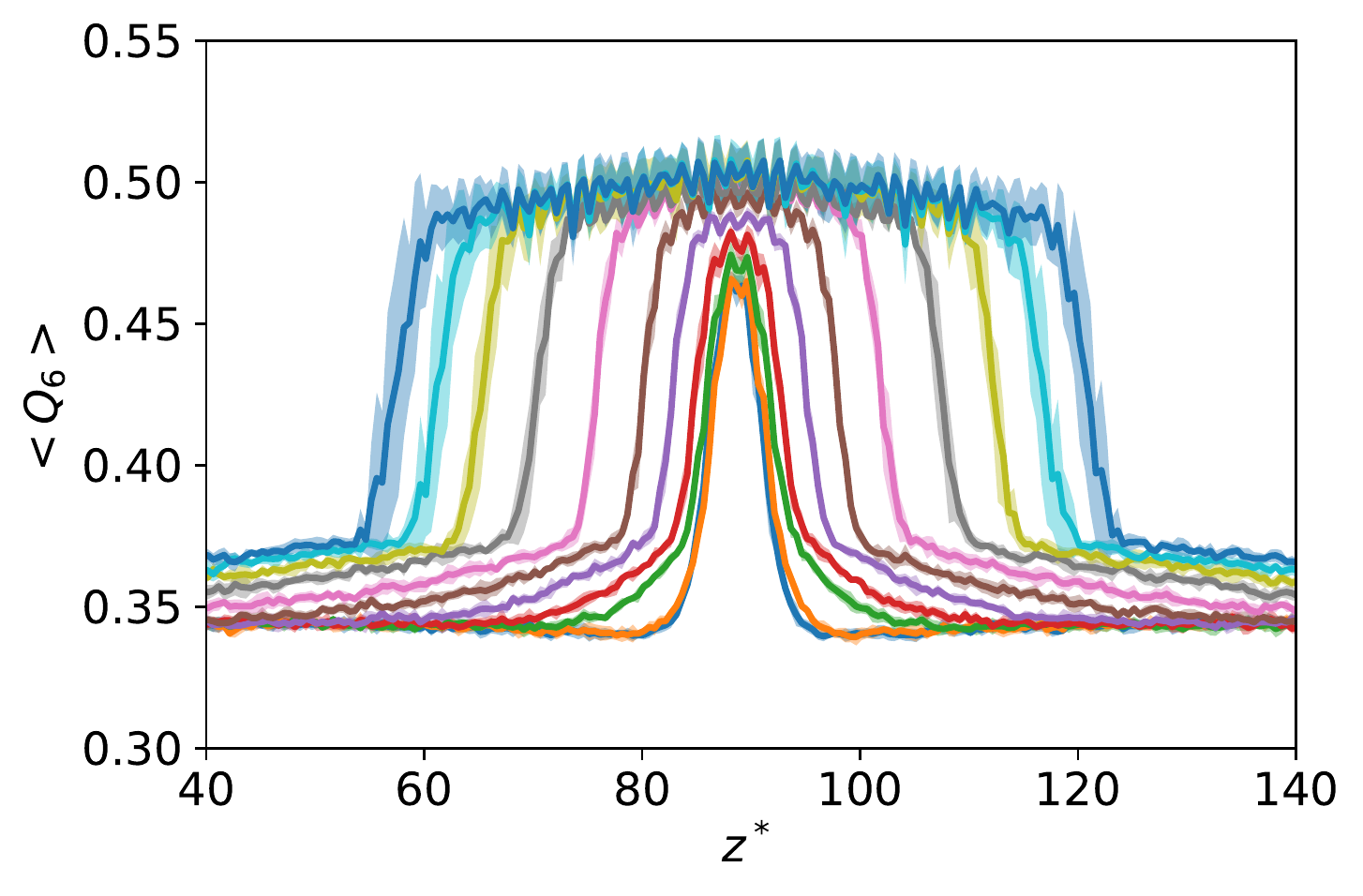}
\end{tabular}
\end{center}
\caption{Transient density, temperature and order parameter $Q_6$ for systems of different sizes, 14$\times$14$\times$56 (left) and  14$\times$14$\times$112 (right) unit cells. The number indicate the corresponding property at a specific time step (1 time step $\equiv$ $\delta t^* = 0.002$). The shadowed regions in each line indicate the standard error, obtained using averages over seven independent simulations. The data for the density at different times (top panels) have been shifted upwards 0.5 units, to show the evolution of the density front.}
\label{fig4}
\end{figure}

\begin{figure}[ht]
\begin{center}
\includegraphics[scale=0.9,angle=0]{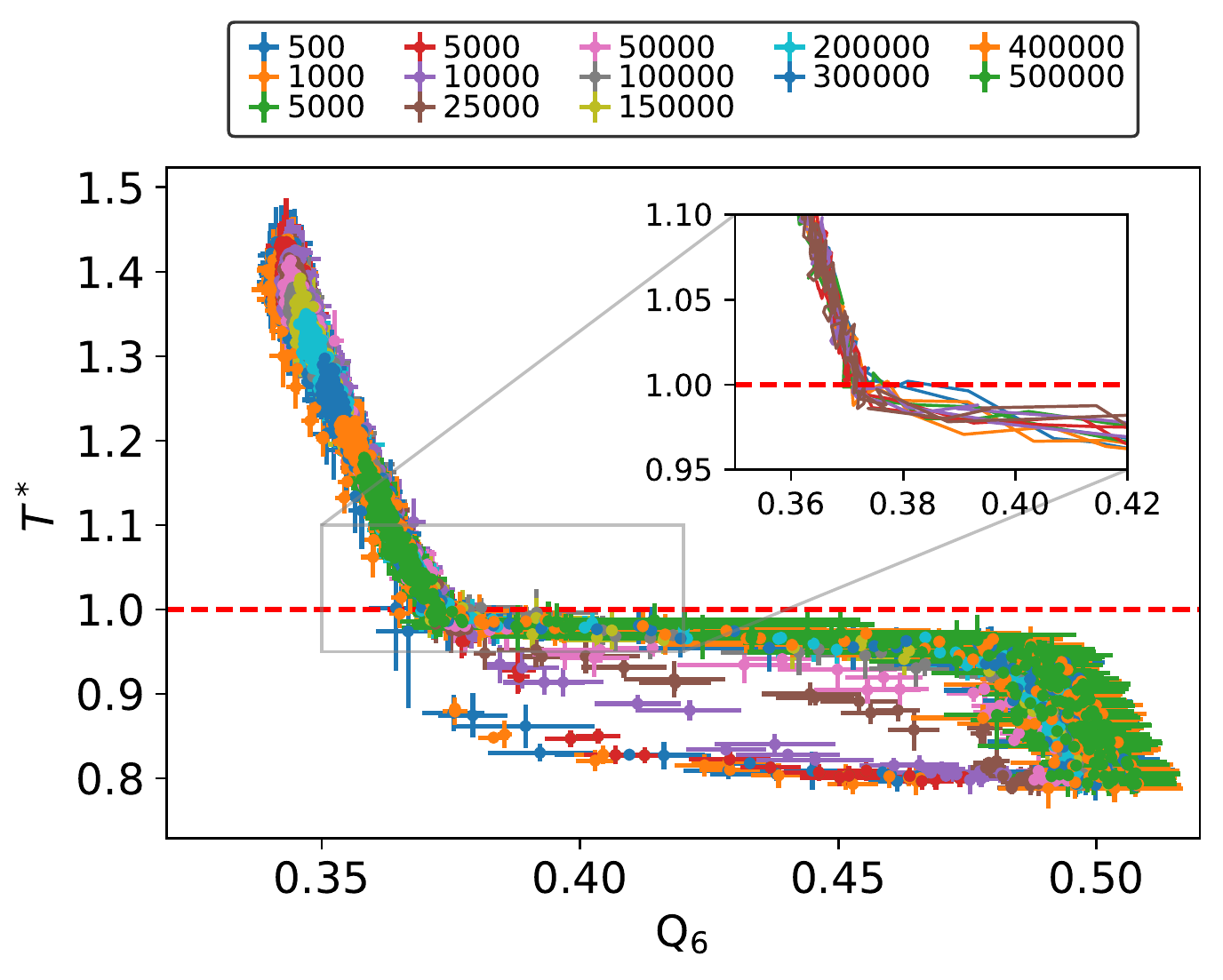}
\end{center}
\caption{Temperature {\it vs} the $Q_6$ local order parameter for systems at different times during the transient freezing process. The inset shows a zoom of the region where the order parameter changes from liquid (left) to solid (right).  The lines in the inset represent the data in the main plot at times in the interval 200-1000 in Lennard-Jones units (corresponding to 1$\times 10^5$-5$\times 10^5$ in simulation times steps). The horizontal dashed lines represent the melting temperature of the crystal for this model at the simulated pressure.  }
\label{TvsQ6}
\end{figure}

We demonstrate in Figure~\ref{fig4} the evolution of the temperature, density and order parameter profiles associated with the freezing process. Initially, the temperature is homogeneous in the liquid phase and it features  a large drop at the center of the simulation box, which corresponds to the region with the thermostat. The solid phase can be readily identified by inspecting the density, $\rho^*$, and the order parameter profiles, $Q_6$ (see Figures~\ref{fig4}(a),(b) and \ref{fig4}(e),(f), respectively). The order parameter is particularly helpful to identity the transition from the liquid to the solid phase, as it features a sharp drop across the liquid-solid interface. The maximum value for the order parameter, $Q_6 \approx 0.5$, is close to the theoretical value for a perfect fcc lattice (using 12 neighbors) $Q_6=0.575$~\cite{mickel2013}. The deviation indicates lattice disorder with respect to a perfect lattice. The disorder is compatible  with the thermal fluctuations present in our simulations. The caveats of using the Q$_6$ parameter to identify the fcc structure have been discussed before \cite{mickel2013}. We did not attempt other approaches, since the order parameter employed here allows us to distinguish the liquid and solid phases very well, making it possible to identify the location of the liquid-solid interface and its time evolution precisely (see Figure~\ref{fig4}).   We tested this idea further by generating correlation plots of the temperature, $T^*$, and the order parameter, $Q_6$ (see Figure~\ref{TvsQ6}). After a transition period $\sim$ 5$\times 10^4$ steps ($t^*=100$ in Lennard-Jones units, or $\sim 200$\,ps)  the correlation $T$ vs. $Q_6$ lines follow a master curve (see Figure~\ref{TvsQ6}), which features an abrupt change in slope at a temperature close to the freezing temperature for the Lennard-Jones model, $T^* \sim 1$  (see ref.\cite{mastny}).  Hence, the correlation plot $T$ vs. $Q_6$ can be used to estimate the freezing temperature ($T_f^*$) of our model (see inset in Figure~\ref{TvsQ6}). This is useful since, unlike in the transient melting process, the temperature profiles do no feature a clear discontinuity in the derivative at $T_f^*$ (\textit{cf.} temperature profiles in Figure~\ref{fig4} and those reported in reference \cite{FontBresme2018}).

The deviations of the simulation data from the $T/Q_6$ master curve in Figure~\ref{TvsQ6} indicate that the results at very short times do not conform to the expected freezing process. At short times, the freezing front is still fairly close to the thermostatting region, and the transition in $Q_6$ between the solid and liquid phases is not as sharp as the one observed at long times, when the interfacial freezing front is well-formed and has moved significantly away from the thermostatting region (c.f. $Q_6$ profiles for $10^4$ and $5\times 10^4$ in Figure~\ref{TvsQ6}). We expect that a well-defined liquid-solid interface moving transiently should conform the  master curve $T/Q_6$ , as observed at long times  $t^* > 100$  ($5\times 10^4$ simulation steps). We have ruled out that the deviations from the master curve at short times are caused by finite size effects related to the thermostat (see Supplementary Information). Hence, we interpret the deviations as a limitation of the order parameter $Q_6$ to capture the very early stages of the structure formation.

\subsection{Comparing theoretical models and MD simulations} 
\label{3.2}

\begin{figure*}[]
\centering
\includegraphics[scale=.5]{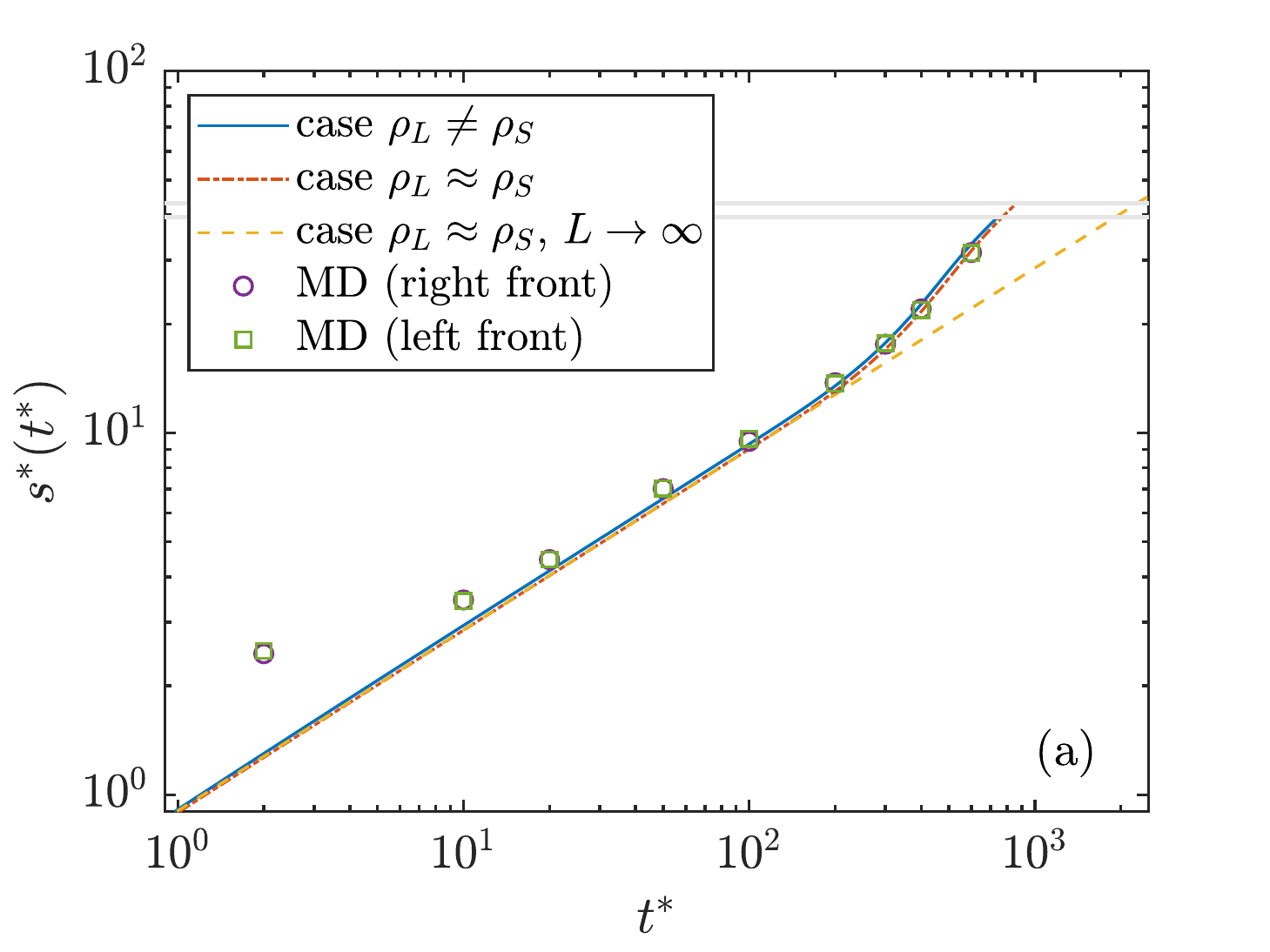}\includegraphics[scale=.5]{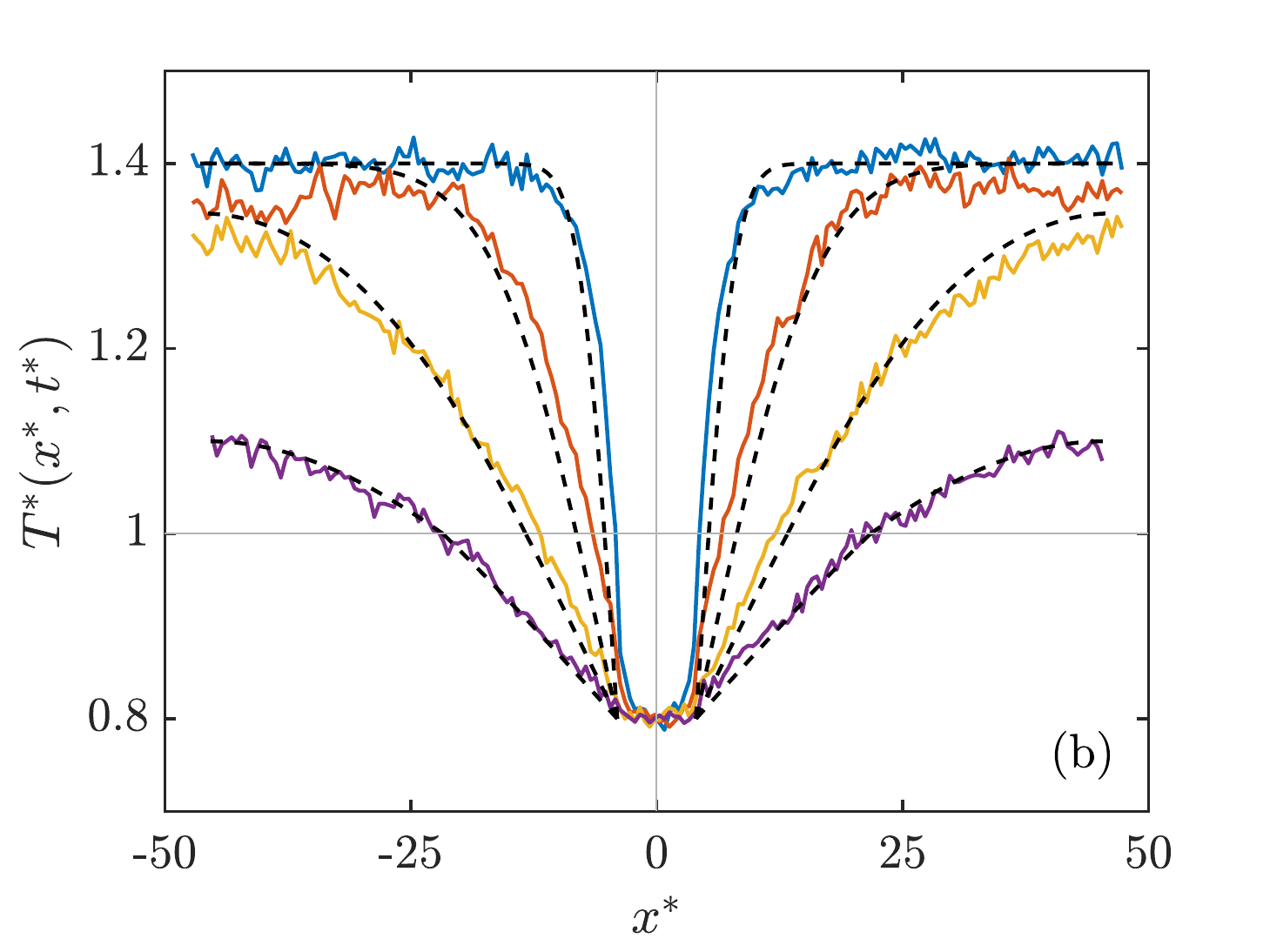}
\includegraphics[scale=.5]{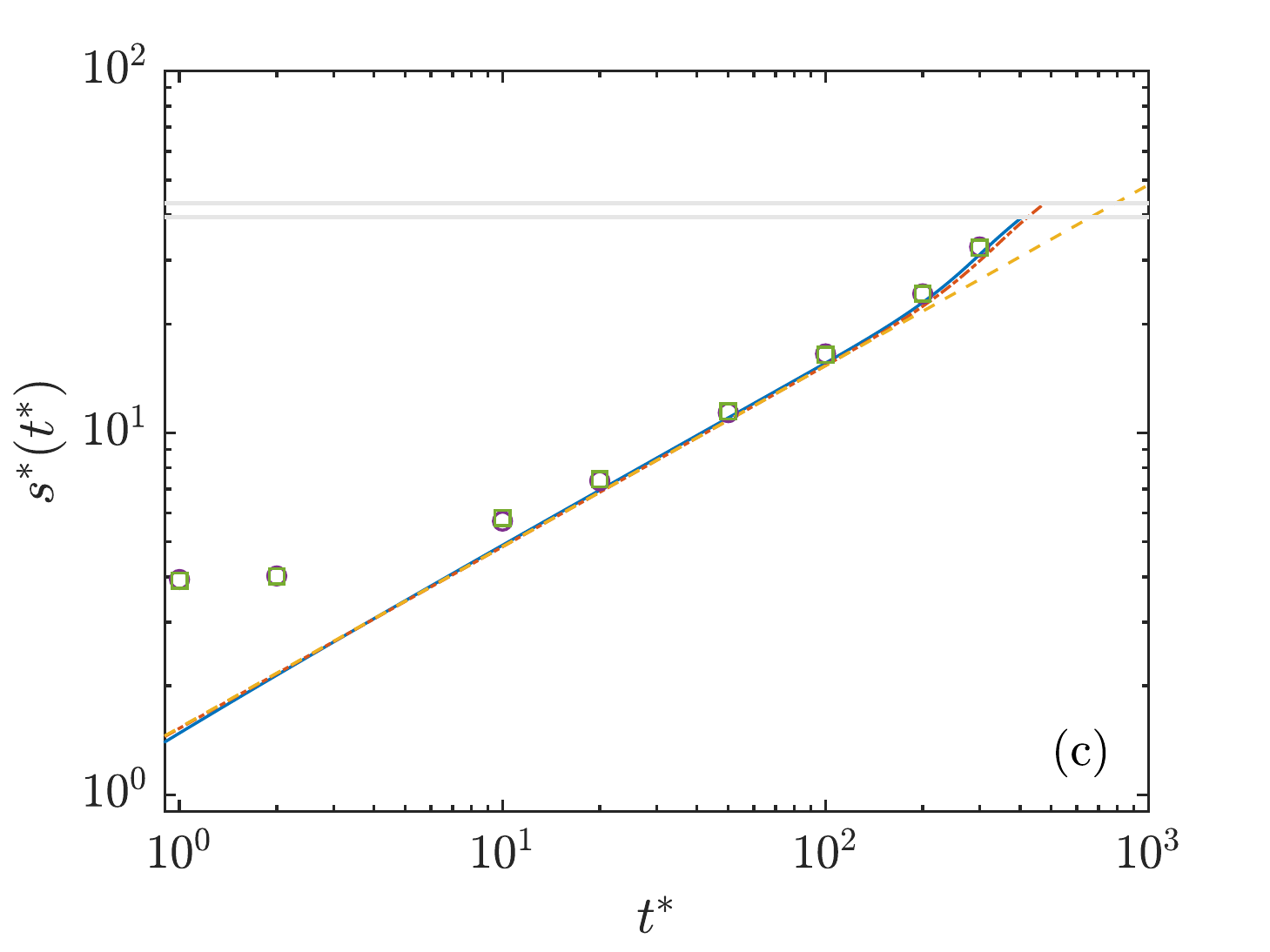}\includegraphics[scale=.5]{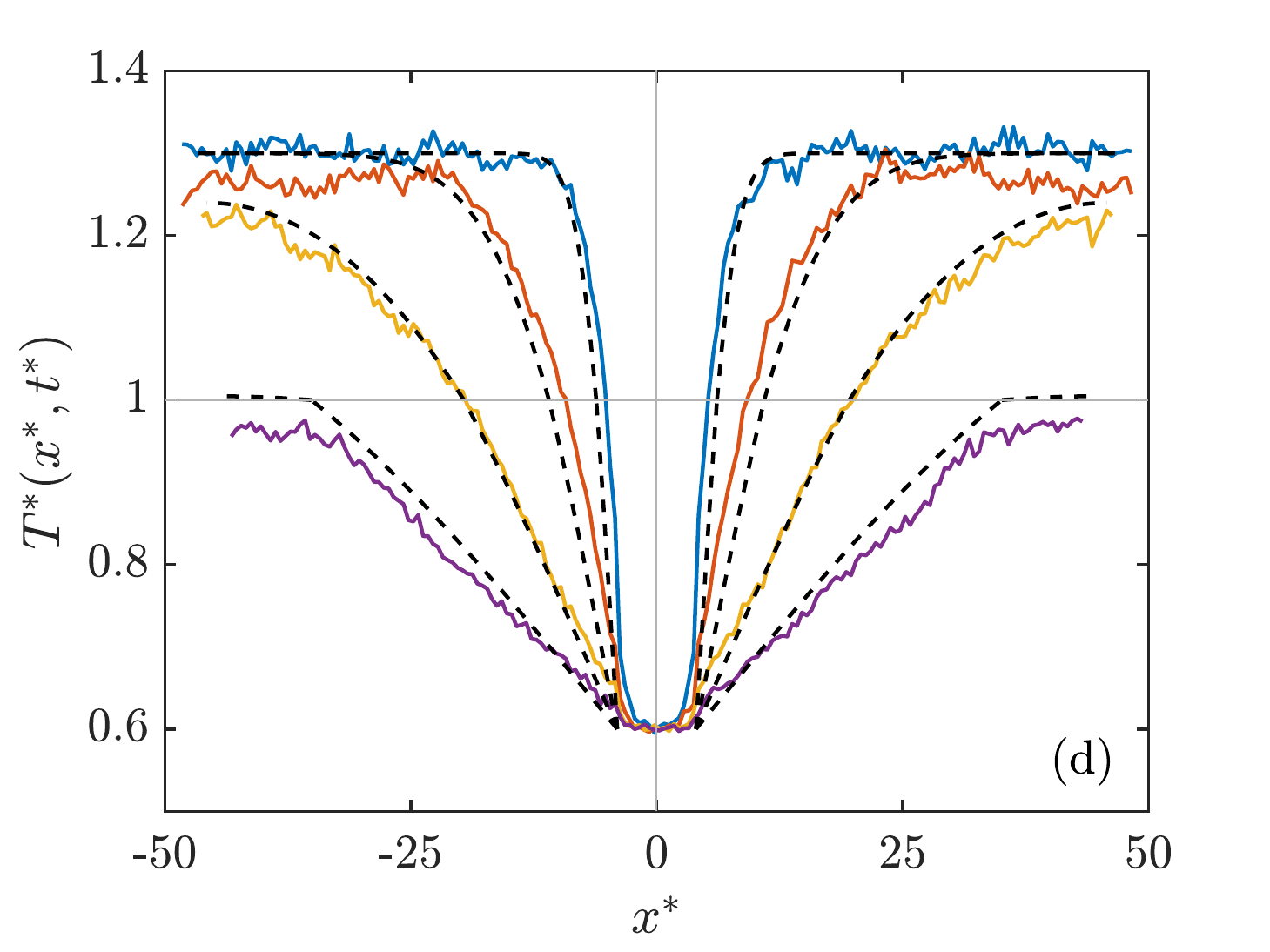}
\includegraphics[scale=.5]{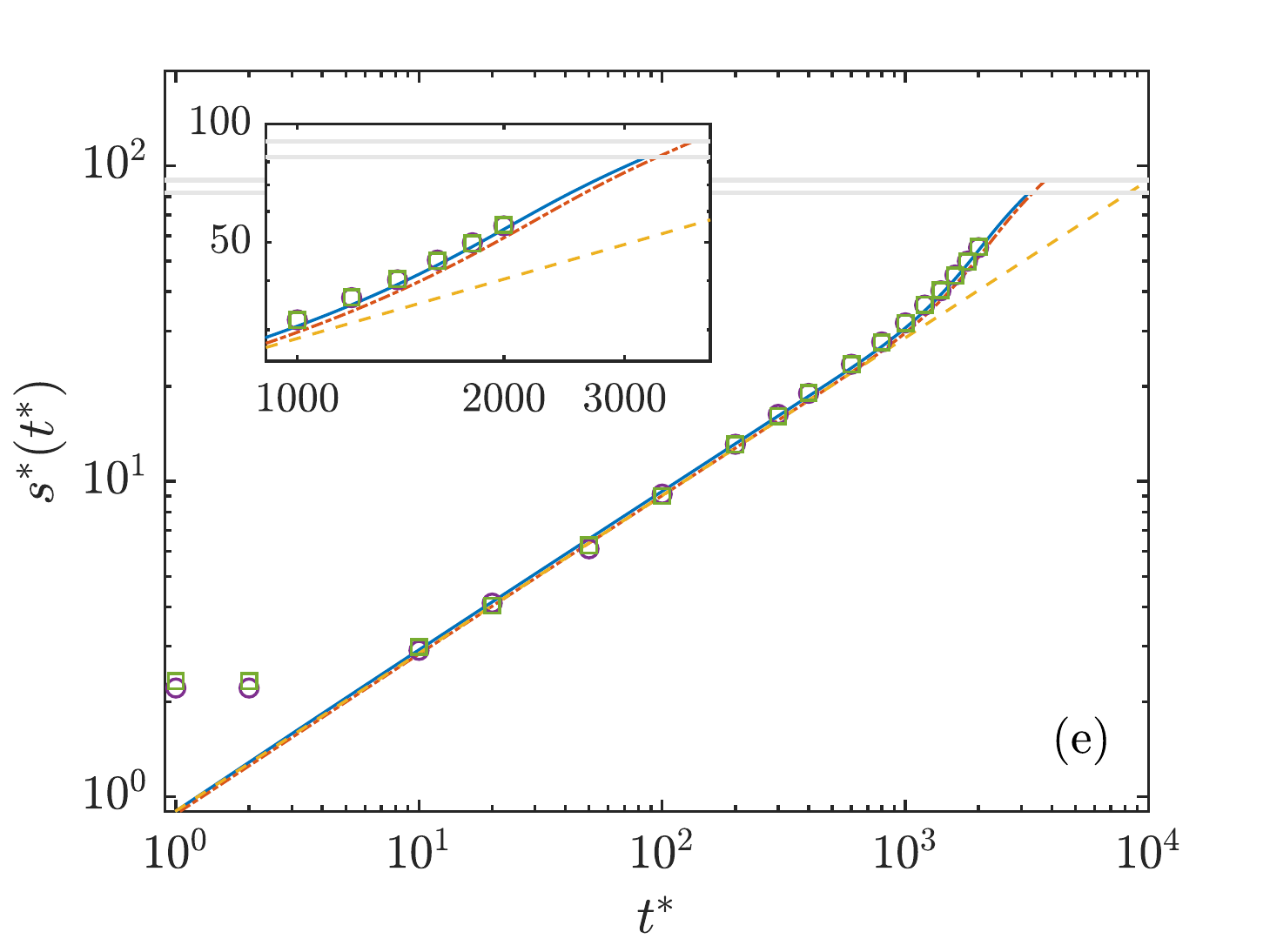}\includegraphics[scale=.5]{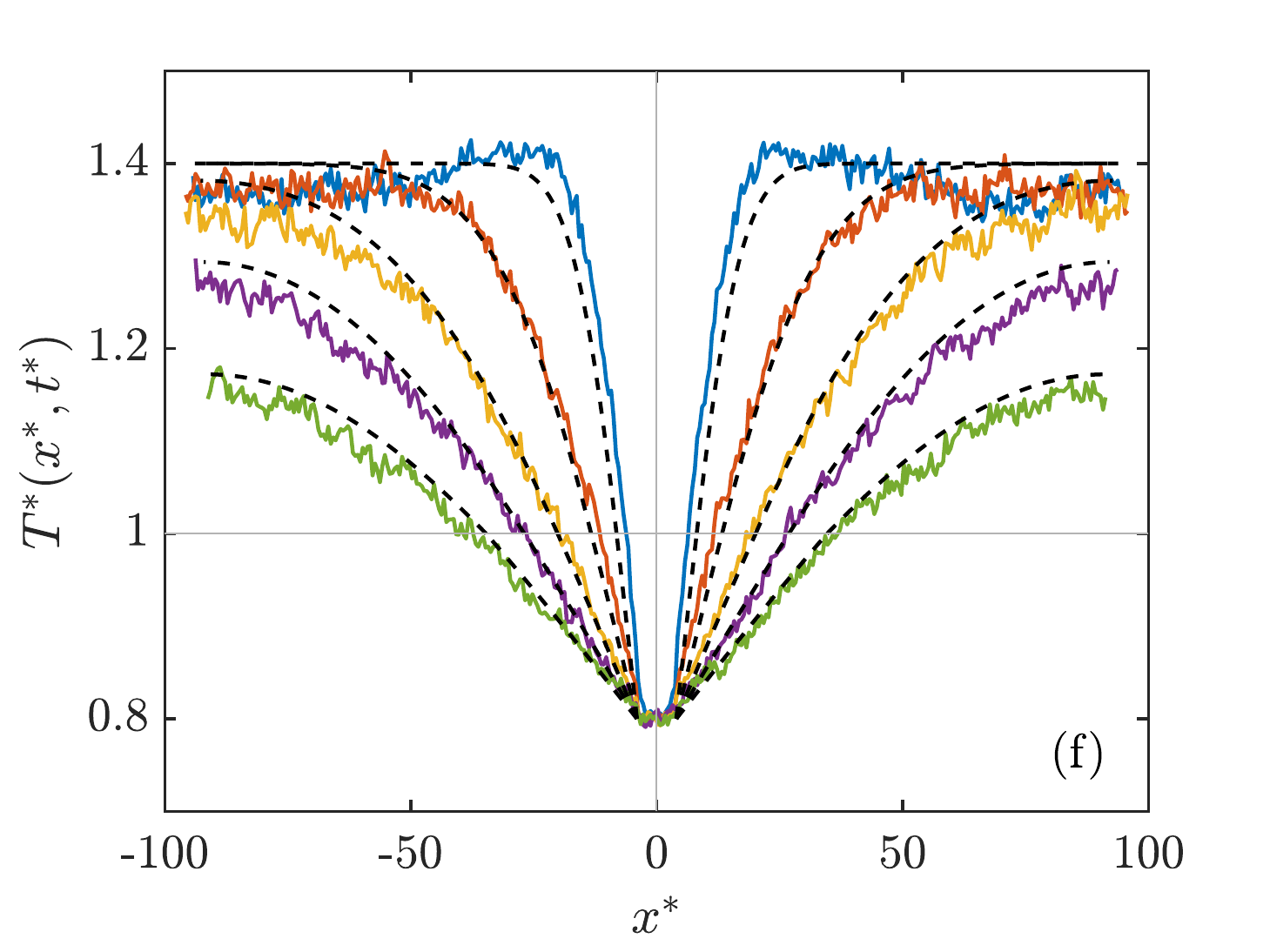} 
\caption{Time evolution of the freezing front (left panels) and the corresponding temperature profiles (right panels). The symbols and lines in the left panel represent the simulation and theoretical results, respectively. The solid lines in the temperature profiles represent the NEMD simulations, and the dashed lines the theoretical prediction for the case $\rho_s\neq\rho_s$.  Panels (a)-(b) represent the results for the system 14$\times$14$\times$56 with $T_0^*=1.4$ and $T_c^*=0.8$. Panels (c)-(d) for system 14$\times$14$\times$56 with $T_0^*=0.6$ and $T_c^*=1.3$, and panels (e)-(f) for the system 14$\times$14$\times$112 with $T_0^*=0.8$ and $T_c^*=1.4$. The temperature profiles in (b) and (d) correspond to  $t^* = 2$, $t^* =20$, $t^* =100$ and $t^* =300$, and in (f) to $t^* =20$, $t^* =100$, $t^* =300$, $t^* =600$ and $t^* =1000$. The inset in (e) shows the late stage of the solidification process. The horizontal grey lines in (a), (c), (e) indicate the final position of the solidification front for the case $\rho_L\approx\rho_S$ (upper line) and $\rho_L\neq\rho_S$ (lower line).  
}
\label{fig-fronts}
\end{figure*}

In Figure~\ref{fig-fronts} we show the evolution of the solid-liquid interface and the temperature profiles predicted by the continuum model \eqref{eq1}-\eqref{eq4}, along with the corresponding simulation data. Panels (a)-(b) and (c)-(d) correspond to the small system (14$\times$14$\times$56), with cooling conditions $T_c^*=0.8$, $T_0^*=1.4$ and $T_c^*=0.6$, $T_0^*=1.3$, respectively, and panels (e)-(f) correspond to the large system (14$\times$14$\times$112) with cooling conditions $T_c^*=0.8$, $T_0^*=1.4$. The solid, dashed and dash-dotted lines in the plots of $s^*(t)$ correspond to the prediction of the continuum model using three different solutions. The dashed line represents the case where the size of the system is assumed to be semi-infinite ({\it i.e.}, $L\rightarrow\infty$) and the densities of the solid and the liquid to be approximately equal ($\rho_l\approx\rho_s$). Under these assumptions the model \eqref{eq1}-\eqref{eq4} has the exact solution \eqref{eq5}-\eqref{eq8}, known as the Neumann solution. The dashed-dotted line corresponds to the case where the size of the system is considered finite and equal to the initial size of the system, but the assumption $\rho_l\approx\rho_s$ is maintained. The most general solution is represented by the solid line, which takes into account that the   densities of the liquid and solid  phases are different ($\rho_l\neq\rho_s$), and that the simulation box  shrinks with time due to the growth of the solid phase ({\it i.e}, $L=L(t)$). In the last two cases, where the size of the domain is assumed to be finite, the theoretical model does not have an exact solution: the results shown correspond to the numerical solution using finite differences. 

The results shown in Figures \ref{fig-fronts}(a),(b) and \ref{fig-fronts}(c)-(d) correspond to the simulations of the small system subject to the cooling conditions $T_c^*=0.8$, $T_0^*=1.4$ and $T_c^*=0.6$, $T_0^*=1.3$, respectively. Comparison of panels (a) and (c) reveals that in the case $T_c^*=0.8$, $T_0^*=1.4$ the interface moves slower than the case $T_c^*=0.6$, $T_0^*=1.3$ and the solidification processes are  completed around $t^*\approx730$ ($730\,\tau=1.58$\,ns) and $t^*\approx400$ ($400\,\tau=0.87$\,ns), respectively. In our simulations, the driving force for solidification is the thermostat temperature $T_c^*$, or the temperature difference $T_f^*-T_c^*$, so the colder the temperature $T_c^*$ the stronger the driving force and the faster the solidification. The faster solidification observed in $T_c^*=0.6$ is also consistent with the system being initially at a cooler temperature ($T_0^*=1.3$). A system at an initially warmer temperature will take longer to solidify since more energy is required to cool the system to the corresponding freezing temperature.

The difference in $T_c^*$ and $T_0^*$ for the results in Figures \ref{fig-fronts}(a) and \ref{fig-fronts}(c) has little impact on the time required for the simulations to converge to the theoretical prediction. In both cases the convergence is observed at $t^*\approx 20$ ($20\,\tau=43.4$\,ps). The main difference between the two cases is the position of the freezing front. In~\ref{fig-fronts}(a) the position of the front is $s^*\approx4.47$ ($4.47\sigma=1.5$\,nm) while in \ref{fig-fronts}(c) is $s^*\approx7.43$ ($7.43\sigma=2.53$\,nm).

The results in Figures \ref{fig-fronts}(e),(f) correspond to a system twice as large as the system from Figures \ref{fig-fronts}(a),(b) but using identical cooling conditions, $T_c^*=0.8$, $T_0^*=1.4$. In this case, simulation and theory converge earlier than in the smaller systems, and the agreement is satisfactory for times $t^*>10$ ($10\,\tau=21.7$\,ps) when the position of the solid-liquid interface is around $s^*\approx2.99$ ($2.99\sigma=1.02$\,nm). In this case, the solidification is completed at $t^*\approx3220$ ($3220\,\tau=6.99$\,ns). The deviation between theory and simulation at shorter times can be interpreted in terms of the order parameter. As shown in Figure~\ref{TvsQ6}, a well-defined sharp interface develops between $t^*=10$ to $t^*=100$ (time steps $5\times10^3$ to $5\times10^4$). At these times, the transition in $Q_6$ begins at $T^*\approx1.0$. Below $t^*=10$ the temperature at which the transition in $Q_6$ begins does not occur at $T^*\approx1.0$, it rather happens around $T^*\approx0.85$ (see time steps $5\times10^2$ and $1\times10^3$). It is therefore expected that the continuum model cannot describe the physics for $t^*<10$ since a key assumption of the model is that at the interface the temperature is $T^*=1$. 

Figure~\ref{fig-fronts}(a) shows that up to $t^* \sim 200$, the Newman solution and the two solutions that account for the finite size of the box agree well with each other, and follow the characteristic $s^*\propto\sqrt{t^*}$ functionality from the Neumann solution. Beyond this time, the impact of the boundary conditions is noticeable, as the speed of freezing increases due to the finite size of the box. The MD simulations depart from the $\sqrt{t}$ trend at the same time as the two solutions of the finite system, and they follow the same behaviour until the completion of the freezing process. The discussion above applies to the rest of the system sizes and cooling conditions (see Figure~\ref{fig-fronts}(c) and (e)) investigated here. 

The incorporation of different densities for the liquid and the solid phases does not introduce  noticeable differences in the propagation of the front for most of the freezing process. However, near the completion of the freezing, the solution with the condition $\rho_s\neq\rho_l$  predicts slightly higher rates  than the solution assuming  $\rho_s\approx\rho_l$ (see, horizontal lines indicating the completion of the freezing process in Figure~\ref{fig-fronts}(a),(c),(e)). The inset in Figure~\ref{fig-fronts}(e) reveals a slightly better agreement between simulation and theory, when the theoretical solution accounts for the different densities of liquid and solid phases. Hence, while the differences in density are small ($\approx 8.67\%$), the influence of the density is still noticeable. For instance, the solidification times for case $\rho_s\approx\rho_l$ and case $\rho_s\neq\rho_l$ in Fig.~\ref{fig-fronts}(e) are $t^*\approx3770$ and $t^*\approx3220$, respectively, resulting in a difference of 14.59\%. Indeed, the common assumption $\rho_s\approx\rho_l$ allows to eliminate the advection term in \eqref{eq2} and the term $\propto (ds/dt)^2$ in \eqref{eq4}, which results in a simpler model while as we have demonstrated, it underestimates slightly the freezing rate. 

We show in Figures~7(b),(d),(f) the time dependent temperature profiles obtained from the simulations and the theoretical prediction for the most general case ($\rho_S\neq\rho_L$). The continuum model describes accurately the temperature relaxation, predicting profiles in good agreement with the simulation results. 

\subsection{Calculation of the enthalpy of freezing from the MD simulation data using the continuum model}
\label{3.3}

The agreement between simulation and theory discussed above can be exploited to extract the enthalpy of freezing from the analysis of the transient non-equilibrium simulations. This provides a route to circumvent the need for additional computations involving equilibrium simulations at coexistence conditions, which require precise knowledge of coexistence densities and pressures. Furthermore, we can use the estimate as a consistency check for the theoretical approach.

Our approach exploits the agreement between the Neumann solution \eqref{eq5}-\eqref{eq8} and the NEMD simulation results, which was found to be excellent after the initial transient regime 
(see Figures~\ref{fig-fronts}(a),(c),(e)). The result showed that the evolution of the freezing front can be described very accurately using
the expression $s(t) = 2\lambda \sqrt{\alpha_s\, t}$. We use this feature to extract $\Delta H_f$ from the equation that links $\lambda$ to $\Delta H_f$ in the Neumann solution.

Given the exact functional form of $s(t)$ is known, we only need to find a function $\tilde{s}(t) = 2\tilde{\lambda}\sqrt{\alpha_s t}$ that fits the NEMD data in the region following the power law $\propto t^{1/2}$. The parameter $\tilde{\lambda}$ that best fits the data can be found by solving the least squares minimization problem
\begin{equation}\label{lamfit}
\tilde{\lambda}=\arg\min_{\lambda_*\in\mathbb{R}^+}\,\sum_{i=1}^{N} \left[y_i-f(t_i,\lambda_*)\right]^2\,,
\end{equation}
where $f(t_i,\lambda_*)=\ln(2\sqrt{\alpha_s}\lambda_*) + \frac{1}{2}\,\ln(t_i)$, $y_i$ is the logarithmic transform of the front position from the NEMD simulations (in dimensional units) and $N$ the number of data points. 
The resulting minimization problem is equivalent to performing a linear regression where the slope is known ($=1/2$) and the only unknown is the intercept (represented by $\ln(2\sqrt{\alpha_s}\lambda_*)$). In this case the problem has the following analytical solution 
\begin{equation}\label{lamfitsol}
\tilde{\lambda} = \frac{1}{2\sqrt{\alpha_s}}\exp 
\left\{
\sum_{i=1}^{N} \left[ y_i - \frac{1}{2}\ln(t_i) \right]
\right\}\,.
\end{equation}
Finally, the value for the enthalpy of freezing is found using
\begin{equation}\label{DHfit}
\Delta \widetilde{H}_{f}= \frac{1}{\tilde{\lambda}\sqrt{\pi}}\left[\frac{c_s(T_f-T_c)}{\exp(\tilde{\lambda}^2)\text{erf}(\tilde{\lambda})} - \frac{c_l(T_0-T_f)}{\sqrt{\alpha_s/\alpha_l}\exp(\alpha_s\tilde{\lambda}^2/\alpha_l)\text{erfc}\left(\tilde{\lambda}\sqrt{\alpha_s/\alpha_l}\right)}\right]\,.
\end{equation}

\begin{figure}[ht]
\centering
\includegraphics[scale=.5,angle=0]{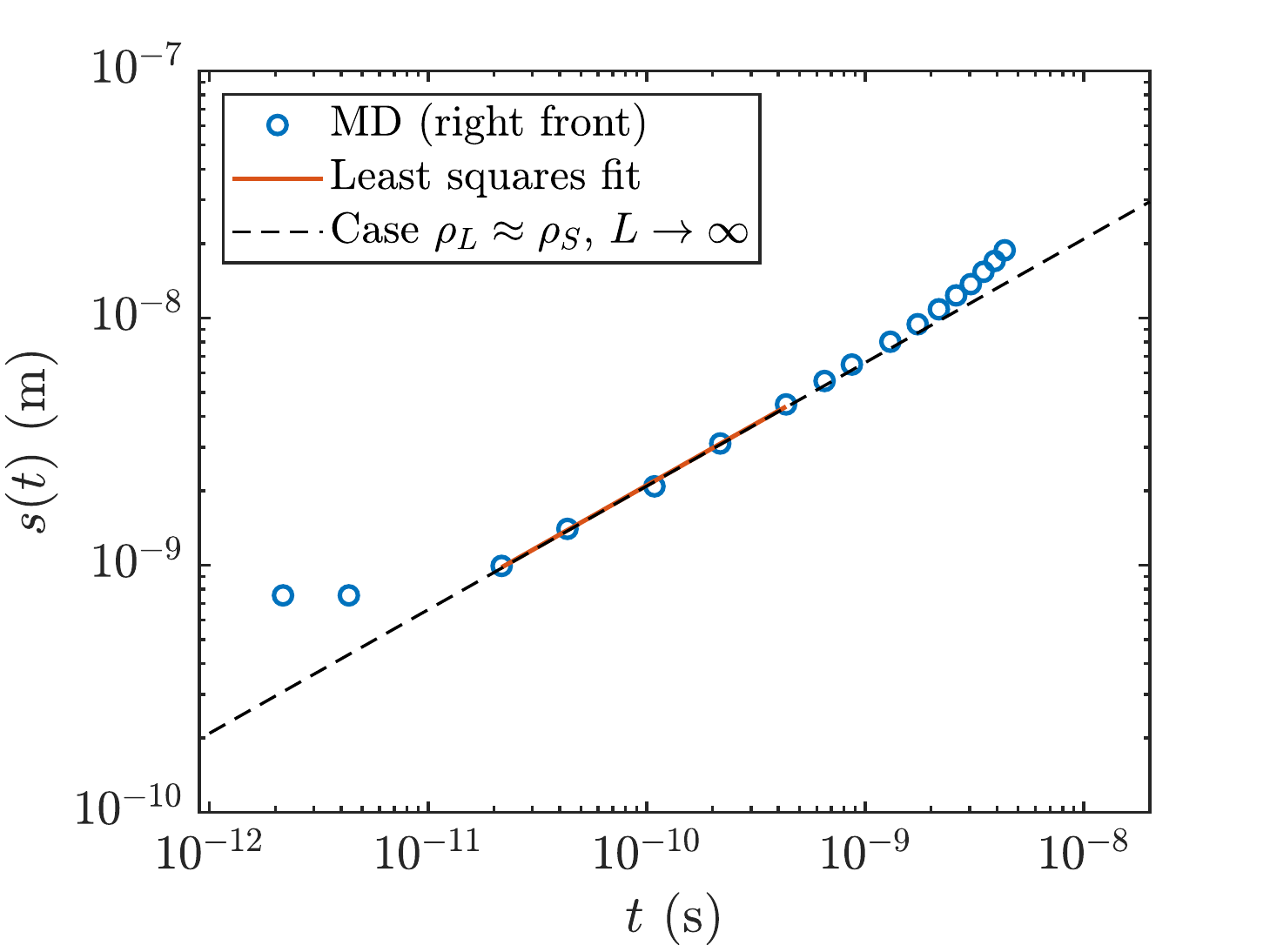}
\caption{Prediction of the evolution of the front by means of the least squares fit $\tilde{s}(t)$ (solid line), along with the MD simulation data for the system 112 (circles) and the corresponding analytical solution of the semi-infinite model $s(t)$ (dashed line).} 
\label{fig_enthalpy}
\end{figure}

To test this procedure we choose the NEMD data for the larger system (14$\times$14$\times$112), and selected the first five data points that followed the $\sqrt{t}$ functionality  (see Figure~\ref{fig-fronts}(e)). From formula \eqref{lamfit} one gets $\tilde{\lambda} = 0.2622$ and using \eqref{DHfit},  $\Delta \widetilde{H}_f = 30956$\,J$\cdot$kg$^{-1}$. These values are very close to the simulated value $\Delta H_{f} = 32492$\,J$\cdot$kg$^{-1}$ (see ref.~\cite{FontBresme2018}) and the value $\lambda = 0.2604$ obtained by solving \eqref{eq8}. In Figure~\ref{fig_enthalpy} we show the simulation data, the fitted function $\tilde{s}(t) = 2\tilde{\lambda}\sqrt{\alpha_s t}$ (solid line) and the prediction from the Neumann solution (dashed line). The agreement is excellent, with the fitted function $\tilde{s}(t)$ and the Neumann solution $s(t)$ being virtually indistinguishable. 

\subsection{Rate of solidification under supercooling conditions} 
\label{3.4}

We have presented above a full analysis of the transient freezing process under a thermal gradient. One question of fundamental and practical interest, {\it e.g.} in material processing, is how the solidification rate compares with the one that might be obtained from a homogeneous freezing process, where the whole system is maintained at the same temperature. To address this point, we performed transient freezing simulations at equilibrium conditions, by quenching the temperature of the entire system to temperatures $T^*=0.8$ and $T^*=0.85$, and performed the simulations by coupling the whole system to this temperature via a thermostat (see discussion in Methods section). In homogeneous freezing the liquid is maintained in a supercooled state, and the velocity of the solidification front is constant in time. This velocity was theoretically described for the first time in the seminal works of Wilson and Frenkel~\cite{wilson,frenkel} 
\begin{equation}\label{WF}
v(T)=\frac{D(T) a}{l^2} f_0 \left(1- \exp(\Delta \mu/k_B T) \right)\,,
\end{equation}
where $D(T)$ is the self diffusion coefficient, $\Delta \mu (T) = \mu_s - \mu_l$ is the difference in chemical potential between the crystal phase and the metastable liquid at temperature $T$, and $l$ is the mean free path for the freezing event. The term $D/l^2$ represents a frequency for a ``jump'' towards the crystal of atoms on the liquid layer of thickness $a$ located next to the surface of the crystal~\cite{Broughton1982}. The constant $f_0<1$, accounts for the atomic collisions that do not result in crystallisation. Using the approximation $\Delta \mu \approx -\Delta H_f \Delta T / T$ (valid for values of $T$ very close to $T_f$) to simplify the exponential term, equation \eqref{WF} can be reduced to a linear function in $T$. The resulting expression has been shown to reproduce homogeneous crystallisation data obtained from molecular dynamics at temperatures in the vicinity of $T_f$ \cite{Mendelev}. We are interested in supercooling conditions where the temperatures are substantially lower than the freezing temperature. Hence, we have considered in our analysis the full equation, including explicitly the chemical potentials.

Equation \eqref{WF} was tested by Broughton \et \cite{Broughton1982} for the same type of  Lennard-Jones (LJ) model investigated here. These authors reported measurable rates below the glass transition temperature, and concluded that the LJ model does not feature a  potential energy barrier for  crystallisation in the presence of the liquid-solid interface. They further replaced the diffusion coefficient term in equation \eqref{WF} by the average thermal velocity of the atoms,
\begin{equation}\label{WL_mod}
v=\frac{f_0 a}{l}\sqrt{\frac{3\,k_B\,T}{m}} \left(1- \exp(\Delta \mu/k_B T) \right).
\end{equation}
Broughton \et found that equation \eqref{WL_mod} could reproduce their results accurately in the whole range of temperatures investigated $T\in[0,T_f]$. More recently, equation \eqref{WF} has been shown to be in agreement with MD simulations of solidification in metals if the parameters  $D(T) a f_0/l^2$ are redefined using a constant that depends on the Einstein frequency of the crystal \cite{Sun}. 

\begin{figure}[ht]
\centering
\includegraphics[scale=.5,angle=0]{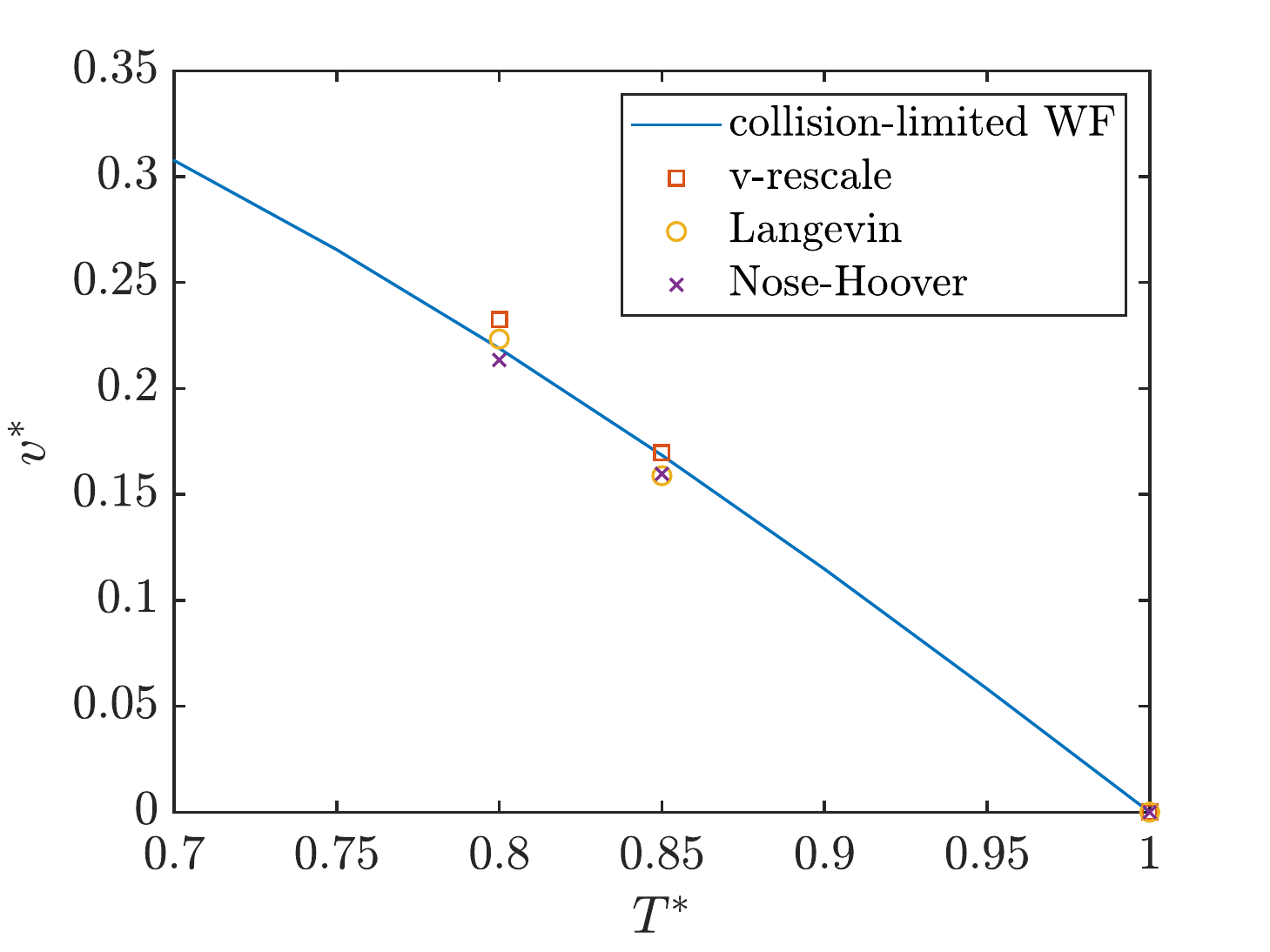}
\caption{Solidification front velocities obtained by MD simulations at two different supercooling temperatures $T^*=0.8$ and $T^*=0.85$ by means of a v-rescale (squares), Langevin (circles) and Nos\'e-Hoover (crosses) thermostats. The solid line represents the prediction by the Wilson-Frenkel theory in the collision-limited regime with $C=0.151$.} 
\label{fig_final}
\end{figure}

In Figure~\ref{fig_final} we show the velocity of the crystallisation front for simulations of the small system (14$\times$14$\times$56), which were performed at supercooling temperatures $T^*=0.8$ and $T^*=0.85$. To ensure that the values of the front velocity are  robust, we performed simulations using three different thermostats (see Methods section for a discussion). The agreement between the velocities obtained with the different thermostats is excellent, showing that our thermostatting approach is not affecting the velocity of the crystallisation front. Along with the simulation results we plot the prediction of \eqref{WL_mod} which gives an estimate of the velocities in the range $T^*\in[0.7,\,1]$, where the parameters $f_0\,a/l$ have been lumped in a single constant $C$ that has been obtained via a least squares fit. We obtained the chemical potential difference by numerical integration of the Gibbs-Helmholtz equation, $\left(\partial \Delta \mu / \partial T \right)_p = (H_{l,m} - H_{s,m}) / T^2$, where $H_{i,m}$ are the molar enthalpies of liquid and solid phases, in a range of temperature spanning  $T^*=[0.7, 1]$.
We find that equation \eqref{WL_mod} fits our simulation results very well. Using the fit, we obtain the values $v^*=0.219$ and $v^*=0.169$ for the temperatures $T^*=0.8$ and $T^*=0.85$, respectively.

To facilitate the comparison with the constant velocities obtained with the homogeneous supercooling approach, we computed the average speed of the crystallisation front  for the three systems presented in Fig.~\ref{fig-fronts}(a), (c) and (e) ({\it i.e}, the cases of solidification with temperature gradients). The average was calculated as $\bar{v}^*=s^*(t_{end}^*)/t_{end}^*$, where $t_{end}^*$ is the solidification time obtained from the solution of the full model ({\it i.e}, case $\rho_l\neq\rho_s$ in Fig.~\ref{fig-fronts}). We found $\bar{v}^* = 0.053$, $\bar{v}^*=0.097$ and $\bar{v}^*=0.026$, respectively. Comparing similar systems, for the small system (14$\times$14$\times$56) solidifying with the thermostat at $T^*=0.8$ (Fig.\ref{fig-fronts}(a)) we obtained $\bar{v}^* = 0.053$ ($\approx8.32$\,m/s) while the homogeneous solidification at the same temperature resulted in $v^*=0.219$ ($\approx 34.37$\,m/s). Therefore, the velocities obtained in supercooled conditions are much faster than in the case of solidification with temperature gradients.

In order to further analyse the difference between solidification rates, we solve the full model now considering the whole liquid already at the freezing temperature by setting $T_0\equiv T_f$ in \eqref{ic} and keeping the thermostat temperature at $T_c=0.8\varepsilon/k_B$. In this case, the temperature of the liquid will remain constant throughout the process ($T_l(x,t)=T_f$) and the velocity will be the maximum velocity that can be achieved, since no extra energy is needed to cool down the liquid (note that we have $T_l(x,t)=T_f$, hence  $\partial T_l/\partial x|_{x=s(t)} =0$ in the Stefan condition). In this case, we obtain the average velocity  $\bar{v}^* = 0.081$ ($\approx 12.71$\,m/s). So, even in the most favourable case for solidification with temperature gradients the velocity will be much lower than the one for homogeneous freezing.

\section{Conclusions}

There have been very few studies exploring the applicability of continuum heat transfer theory based on Fourier law to describe nanoscale heat transfer involving phase change. In this work, we have investigated the accuracy of the continuum theory to describe transient freezing processes in small systems and short time scales. We have tested the solutions of the standard phase change model based on continuum heat transfer theory against non-equilibrium molecular dynamics simulation data. The comparison reveals that differences between theory and simulations only occur at short times, below 50\,ps when the solid-liquid interface has moved a distance smaller than 3\,nm (relative to the edge of the thermostat) in the direction of growth. In particular, we find that for a system with length of 30.27\,nm the time required to observe the onset of the behavior described by the continuum theory (convergence time) is about 40\,ps when 
the solid-liquid interface has travelled a distance of about 1.5\,nm or 2.5\,nm, depending on the cooling conditions. For a larger system of length 60.54\,nm the convergence time is about 20\,ps when the solid-liquid interface has travelled a distance of 1\,nm.

For times longer than the convergence time, the simulated crystal grows at the rate predicted by the theory, {\it i.e.} with the position of the solidification front moving as $\sqrt{t}$. Towards the end of the solidification process the position of the solid-liquid interface deviates from the  $\propto \sqrt{t}$ behavior, accelerating until all liquid is solidified. This behavior is connected to the finite size of the simulation box and it is well captured by the theory, when the finite domain is considered in the solution of the phase change model. We also found that the agreement between theory and simulations improves when the densities of the solid and liquid phases are considered explicitly. Generally we find that the agreement between  simulation and theory is excellent in the regime where the front follows the $\sqrt{t}$ behavior. We use this result to introduce an approach to calculate the latent heat of the solid, from the analysis of the transient freezing process using the exact solution of the continuum model.   

Finally, we have investigated the freezing process at homogeneous conditions, namely without temperature gradients. Such a process can be performed experimentally by quenching the temperature of the whole system (thermostat and liquid) under the solidification temperature, and maintaining the whole system at the selected subfreezing temperature. Hence, the crystal grows into a supercooled liquid with a constant homogeneous temperature. We find that the speed of the freezing front in the homogeneous process is one order of magnitude larger than the rates obtained with the temperature gradients for systems of the same size subject to the same thermostat temperature.

\section*{Acknowledgements}

FB thanks the EPSRC-UK (EP/J003859/1) and The Leverhulme Trust (grant RPG-2018-384) for  financial support and  the Imperial College High Performance Computing Service for providing computational resources. FF acknowledges financial support from the \emph{Juan de la Cierva} programme (grant IJC2018-038463-I) from the Spanish MICINN, from the {\it Obra Social la Caixa} through the programme {\it Recerca en Matem\`{a}tica Col$\cdot$laborativa} and the CERCA Programme of the {\it Generalitat de Catalunya}. 

\section*{Supplementary Information}

\begin{figure}[ht]
\centering
\includegraphics[scale=.75]{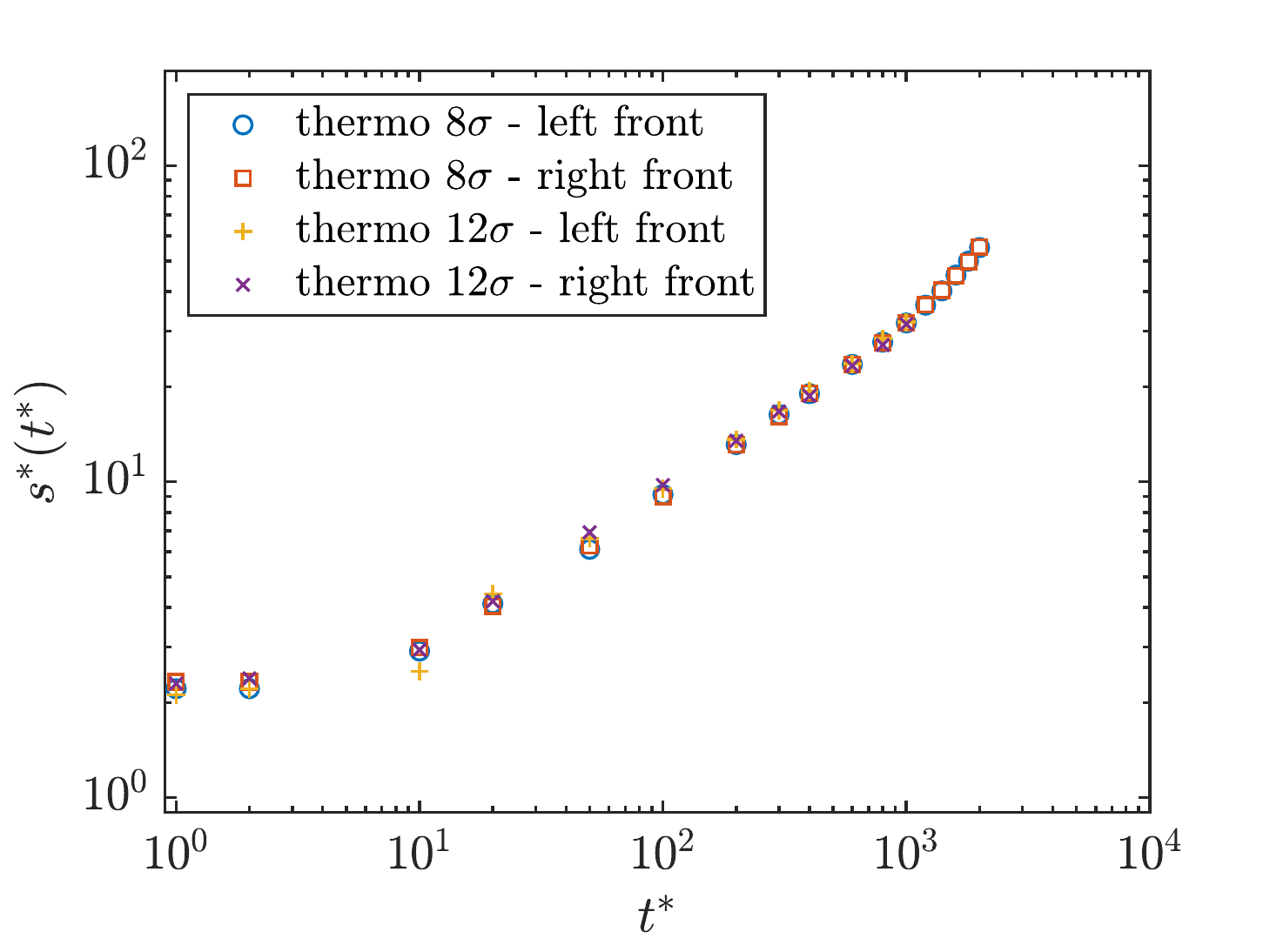} 
\caption{Time evolution of the left and right fronts from the NEMD simulations for the system 14$\times$14$\times$112 with $T_0^*=0.8$ and $T_c^*=1.4$ using a thermostat of size 8$\sigma$ and 12$\sigma$. The $y$-axis shows values relative to the edge of the thermostat. }
\label{figSI}
\end{figure}

In order to rule out any size effect of the thermostat on the simulation results we performed an additional simulation for the large system (14$\times$14$\times$112) with a thermostat of size 12$\sigma$. In Figure~\ref{figSI} we compare the evolution of the left and right fronts using the thermostat of size $8\sigma$ presented in the main text (See Fig.~7(e)) and the evolution of the fronts obtained for the simulation with the thermostat of size 12$\sigma$. The position of the fronts is measured relative to the edge of the thermostat (here and in the main text), which ensures that the comparison between the case 8$\sigma$ and 12$\sigma$ is consistent. The comparison shows that there are no relevant differences in the simulation results using a thermostat of size 8$\sigma$ and 12$\sigma$, and therefore we can rule out any spurious effect connected to the size of the thermostat in our results. 

\bibliographystyle{plain}
\bibliography{solbibfilenew}

\end{document}